\documentclass[floatfix,aps,pra,superscriptaddress,twocolumn,longbibliography]{revtex4-2}

\usepackage[english]{babel}

\usepackage{graphicx}
\usepackage[colorlinks=true, allcolors=blue]{hyperref}
\usepackage{physics}
\usepackage{amsmath,amsfonts,amssymb,amsthm}
\usepackage{enumitem}
\usepackage[utf8]{inputenc}

\usepackage{natbib}
\usepackage{url}
\hypersetup{colorlinks=true, urlcolor=blue, linkcolor=blue, citecolor=blue}

\begin{document}

\title{Interaction enhanced quantum heat engine}

\author{Mohamed Boubakour}
\email{mohamed.boubakour@oist.jp}
\author{Thom\'{a}s Fogarty}
\author{Thomas Busch}
\affiliation{Quantum Systems Unit, Okinawa Institute of Science and Technology Graduate University, Onna, Okinawa 904-0495, Japan}

\date{\today}

\begin{abstract}
We study a minimal quantum Otto heat engine, where the working medium consists of an interacting few-body system in a harmonic trap. This allows us to consider the interaction strength as an additional tunable parameter during the work strokes. %during which the external potential and the interaction are simultaneously changed. W%e show that the efficiency of such a cycle can outperform a similar non-interacting cycle. 
We calculate the figures of merit of this engine as a function of the temperature and show clearly in which parameter regimes the interactions assist in engine performance. We also study the finite time dynamics and the subsequent trade-off between the efficiency and the power, comparing the interaction enhanced cycle with the case where the system remains scale-invariant.     
\end{abstract}

\maketitle

%%%%%%%%%%%
% Article %
%%%%%%%%%%%

\section{Introduction}
\label{Sec:Introduction}
The study of quantum heat engines (QHEs) is a central part of the field of quantum thermodynamics \cite{Sai,Alicki,Kosloff2014,Binder_2018}. They can be used to understand the role that quantum effects play when comparing to classical settings, while at the same time they have implications for the development of quantum technologies.
Usually a QHE will consist of a quantum system as the working medium (WM) to which a conventional thermodynamic cycle (Carnot, Otto etc.) is applied in order to extract work from the heat exchanged with a cold and a hot bath. Historically, most works have addressed single particle systems \cite{Abah2012,Scovil,Geva,Humphrey,Hugel,Rossnagel,Kosloff2017,Kosloff2014,Goswami,Newman,Ono}, however more recently QHE that use interacting systems have attracted more attention \cite{Jaramillo_2016,Chen_2019,Carollo_2020,Myers}. In particular, it is interesting to understand the effect of the interaction on the performance, and to identify parameter regimes in which  cooperative effects due to these interactions allow to outperform single particle QHEs \cite{Fogarty_2020,Halpern,  Uzdin,Vroylandt_2017,Niedenzu_2018, Myers_2022}. However, care must be taken as interactions have also been shown to reduce engine performance \cite{Chen}, and the dynamical control of these systems can be more complex leading to the creation of irreversible excitations \cite{Fogarty_2020,Keller_2020, Hartmann_2020,Beau_2016,Deng_2018,Li_2018,Mossy,Mikkelsen}.

In this work we show that a suitable tuning of the interactions can be used to improve the performance of QHEs when compared to systems with non-interacting working media. For this, we consider interacting bosons confined in a harmonic trapping potential and realize the adiabatic compression and expansion strokes of the Otto cycle through increasing and decreasing the trap frequency. However, we also drive the interactions between two distinct values during these strokes and show that optimal interaction strengths exist that increase the work output and the efficiency when compared to a non-interacting engine. 
 
We also show that the effect of the interaction strongly depends on if one considers distinguishable or indistinguishable particles, and we calculate the efficiency at maximum work output (EMW) showing that the interaction significantly improves this quantity in the low temperature regime. Finally we explore the finite time dynamics of the cycle, finding the optimal operation times of the QHE with interacting working media.

The manuscript is organized as follows: in Sec.~\ref{Sec:OttoCycle} we present the model that describes the WM and the thermodynamic cycle of our QHE. In Sec.~\ref{3}, we study the performance in the adiabatic limit by first looking at the case of non-interacting particles and comparing the performance of distinguishable particles and indistinguishable bosons in Sec.~\ref{3a}. We then consider the case of two interacting particles in Sec.~\ref{3b}, and discuss the performance of an engine with two indistinguishable bosons in Sec.~\ref{3c} and with two distinguishable particles in Sec.~\ref{3d}. We extend our results to the case of three interacting particles in Sec.~\ref{3e}. In Sec.~\ref{4} we study the engine performance with two particles for cycles run in finite time, and finally we conclude in Sec.~\ref{5}.  
 
\section{Quantum Otto heat engine with driven interaction}\label{Sec:OttoCycle}

\begin{figure}[tb]
\includegraphics[width=\linewidth]{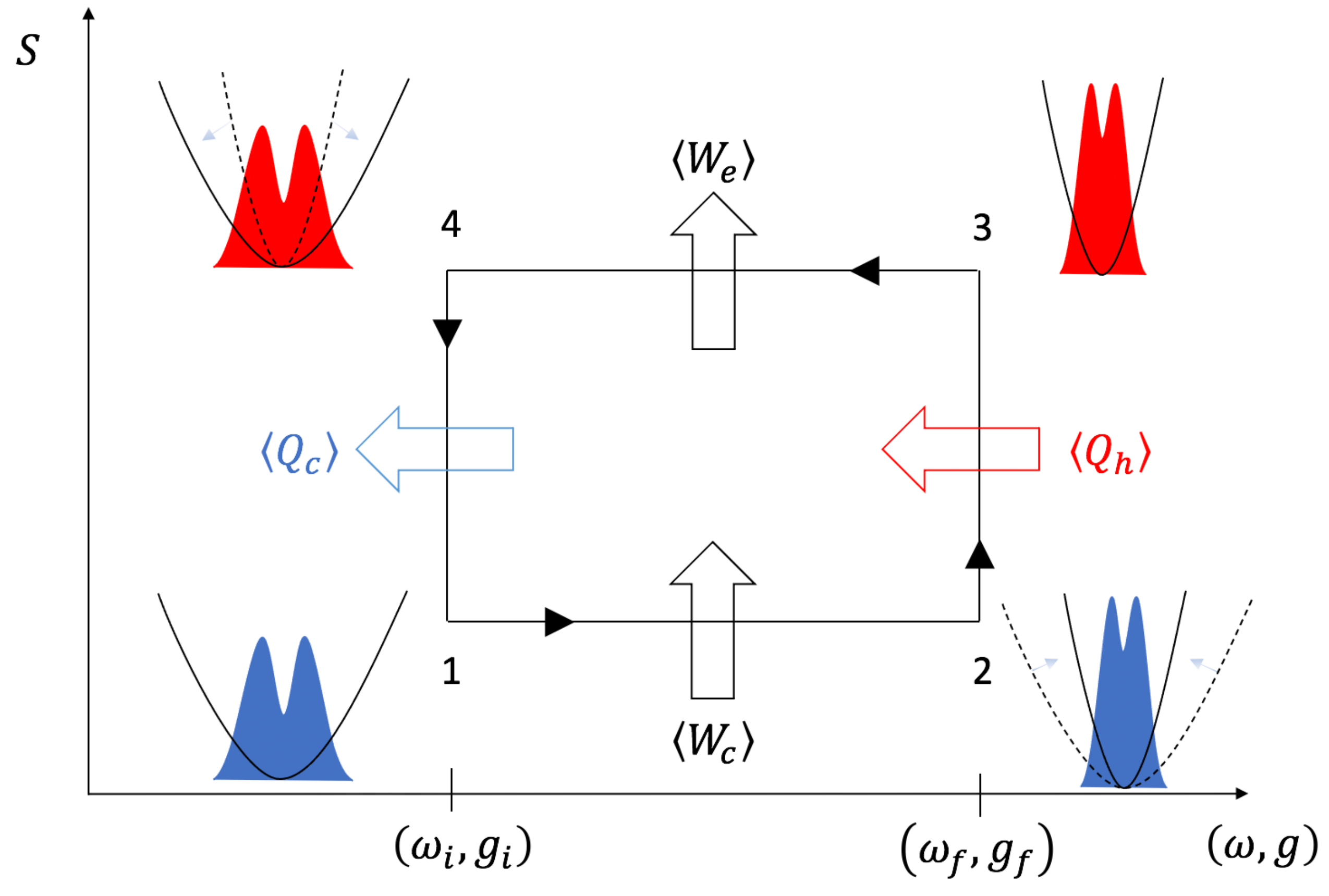}
\caption{Schematic of the heat engine cycle. The y-axis represents the entropy of the WM and the x-axis represent the trap frequency and the interaction.
}

\label{cycle}
\end{figure}

We consider a QHE cycle where the WM is an interacting quantum gas confined to one dimension and trapped in a harmonic potential. The Hamiltonian is given by
\begin{equation}\label{hamiltonian}
    H(\omega,g)=\sum_{n=1}^{N}-\frac{\hbar^{2}}{2m}\frac{\partial^{2}}{\partial x_{n}^2}+\frac{1}{2}m\omega^{2}x_{n}^2 +g\sum_{n<p}\delta(x_{n}-x_{p}),
\end{equation}
where $m$ is the mass of the particles and $\omega$ is the trap frequency. Since we only consider low temperatures, we can approximate the interaction by a point-like potential where $g$ is the 1D interaction strength between the particles. For $N=2$ this Hamiltonian can be analytically solved \cite{Busch98}, however for $N\geq3$ numerical methods are required to find the eigenstates \cite{Weisse_2008,Na_2017}. The engine cycle we explore is similar to a standard Otto cycle except that the adiabatic strokes occur by changing two parameters: the trap frequency $\omega$ and the interaction strength $g$. A schematic is shown in Fig.~\ref{cycle} and the individual strokes are given by
\begin{enumerate}
    \item[] \textit{Adiabatic compression $(1\rightarrow 2)$}: the WM is initially trapped in a harmonic potential with frequency $\omega_{i}$ and at equilibrium with inverse cold temperature $\beta_{c}$. The interaction strength is given by  $g_i$. From there a compression stroke is carried out that performs work on the system by increasing the trap frequency to $\omega_{f}$ and changing the interaction strength to $g_{f}$. The work is given by $W_{c}=\left<H(\omega_{f},g_{f})\right>_{2}-\left<H(\omega_{i},g_{i})\right>_{1}$. 
    
    \item[] \textit{Hot isochore $(2\rightarrow 3)$}: the next stroke increases the temperature of the WM by coupling it to an external hot bath at the inverse temperature $\beta_{h}$ with the control parameters $g_f$ and $\omega_f$ fixed. In equilibrium the heat exchanged during this stroke is given by $Q_{h}=\left<H(\omega_{f},g_{f})\right>_{3}-\left<H(\omega_{f},g_{f})\right>_{2}$.
    
    \item[] \textit{Adiabatic expansion $(3\rightarrow 4)$}: the system is then decoupled from the hot bath and work is extracted from the WM by adiabatically driving the trap frequency and interaction strength back to $\omega_{i}$ and $g_{i}$. The work is given by $W_{e}=\left<H(\omega_{i},g_{i})\right>_{4}-\left<H(\omega_{f},g_{f})\right>_{3}$. 
    
    \item[] \textit{Cold isochore $(4\rightarrow 1)$}: in the last stroke the WM is cooled down by exchanging heat with a cold bath at the inverse temperature $\beta_{c}$. It returns to the initial state and the heat exchanged during this stroke is given by $Q_{c}=\left<H(\omega_{i},g_{i})\right>_{1}-\left<H(\omega_{i},g_{i})\right>_{4}$.
\end{enumerate}

It is worth noting the difference between  adiabatic strokes in the quantum and in the classical regime. While carrying out an adiabatic stroke in a classical setting means that no heat exchange occurs during the process, for quantum systems it refers to the condition that the occupation populations of the eigenstates remain constant. This difference in the definition implies a difference in the time scale of the strokes. In classical heat engines, the WM will be driven quickly in order to prevent the system from relaxing and therefore exchanging heat with the environment, while for QHEs one needs to drive it quasistatically based on the adiabatic theorem. The performance of the engine is characterized by the work output $W=W_{e}+W_{c}$ and the efficiency $\eta=\frac{\abs{W}}{Q_{h}}=1+\frac{Q_{c}}{Q_{h}}$. By convention, we chose the variation of energy to be negative when the WM loses energy, which means that the engine produces extractable work when $W<0$. Like in a conventional quantum heat engine, we choose the trap frequency at the end of the compression to be larger than the initial frequency, $\omega_{f}>\omega_{i}$, however $g_{f}$ can be larger or smaller than $g_{i}$. 

\section{Engine performance in the adiabatic limit}\label{3}

\subsection{Non-interacting limit and statistical influence on the performance}\label{3a}

\begin{figure}
\includegraphics[width=\linewidth]{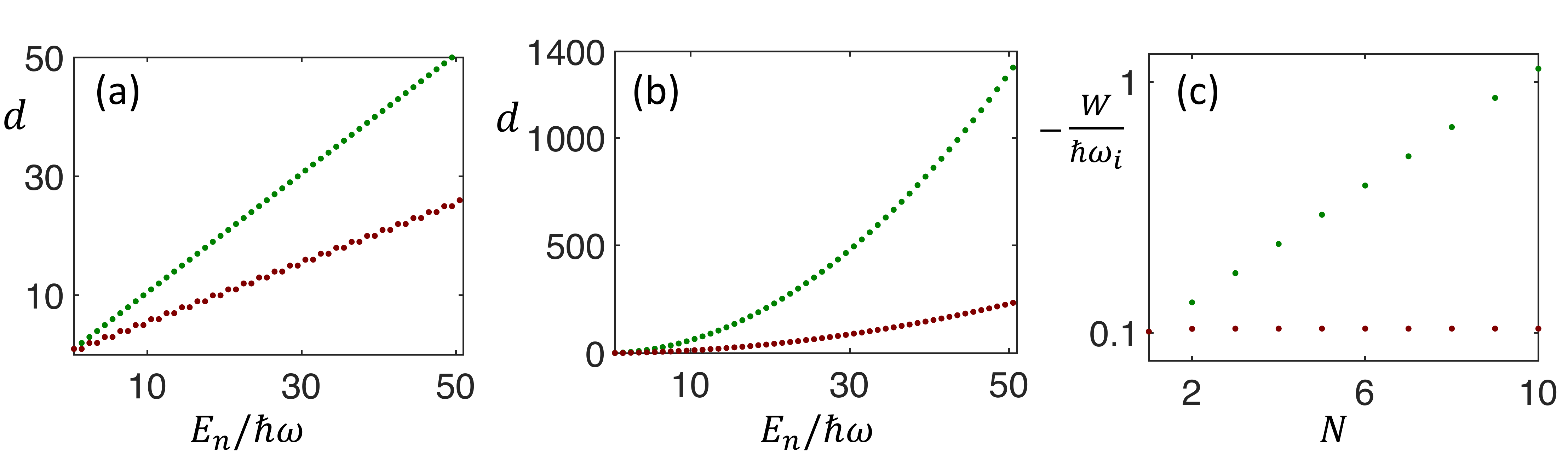}
\caption{Degeneracy of the energy levels for a system of (a) two distinguishable particles (green dots) and two indistinguishable bosons (brown dots) and (b) three distinguishable particles (green dots) and three indistinguishable bosons (brown dots). (c) Work output of the Otto cycle of non-interacting distinguishable particles (green dots) and non-interacting indistinguishable bosons (brown dots) as a function of the number of particles. The compression ratio is $\kappa=\frac{\omega_{i}}{\omega_{f}}=\frac{1}{3}$, the cold inverse temperature is $\beta_{c}=\frac{10}{\hbar\omega_{i}}$ and the hot inverse temperature is $\beta_{h}=\frac{1}{\hbar\omega_{i}}$.}

\label{d_vs_b}
\end{figure}

Before examining the effects of the interactions in the working medium, let us first consider the non-interacting limit ($g_{i}=g_{f}=0$) in order to outline the influence of the statistical properties on the engine performance. Below we consider distinguishable particles and indistinguishable bosons, where their respective statistics leads to a difference in the degeneracy of the energy levels given by 
\begin{equation}
    d(E_{n})=\sum_{n_{1}}...\sum_{n_{N}}\delta_{E_{n},\hbar\omega\left(\sum_{j=1}^{N}n_{j}+\frac{1}{2}\right)},
\end{equation}
where $\delta_{a,b}$ is the Kronecker symbol. We illustrate this in Fig.~\ref{d_vs_b}(a)-(b) for two and three particles systems and unsurprisingly the number of states for a given energy is higher for distinguishable particles. In fact, the gap between these two distributions increases exponentially with the number of particles. The probability for $N$ indistinguishable bosons to be at the same energy is therefore higher than for $N$ distinguishable particles and this increases with the number of particles. In particular, indistinguishable bosons will most likely stay in the ground state and the probability for a boson to transition to an excited state will be small for low temperatures. As a consequence, the performance of an engine realized with non-interacting bosons will be limited in terms of work output in the temperature regimes of our interest (which corresponds, as we will see later, to the temperature regime where interactions lead to interesting behaviors). The respective work output of the Otto-cycle of non-interacting bosons and distinguishable particles as a function of the number of particles is shown in Fig.~\ref{d_vs_b}(c). As expected, the work output for distinguishable particles increases linearly and from physical arguments one can expect the work output for bosons to be sub-linear. However in Fig.~\ref{d_vs_b}(c), one can see that the behaviour is more than sub-linear and it, in fact, reaches a plateau for $N\ge 3$. This means that the mean occupations of the energies for bosons at the hot and cold temperatures become so similar that adding particles only contributes negligibly to the work output. 
Given this drastically different behavior in the non-interacting limit, let us next study  distinguishable particles and indistinguishable bosons in the presence of interactions.

\subsection{Two particles working medium}\label{3b}
We now take account of the interaction by considering first the two particle case ($N=2$). The Hamiltonian can be solved by introducing the center of mass coordinate $X=\frac{x_{1}+x_{2}}{\sqrt{2}}$ and the relative coordinate $x=\frac{x_{1}-x_{2}}{\sqrt{2}}$, which allows one to split the Hamiltonian \eqref{hamiltonian} into two decoupled single particle Hamiltonians $H(\omega,g)=H_{CM}(\omega)+H_{r}(\omega,g)$ with
\begin{align}
    H_{CM}(\omega) &=-\frac{\hbar^{2}}{2m}\frac{\partial^{2}}{\partial X^2}+\frac{1}{2}m\omega^{2}X^2,\\
    H_{r}(\omega,g) &=-\frac{\hbar^{2}}{2m}\frac{\partial^{2}}{\partial x^2}+\frac{1}{2}m\omega^{2}x^2+\frac{g}{\sqrt{2}}\delta(x).
\end{align}
The eigenstates are thus given by the two particle states $\ket{n,\nu}$, where $\ket{n}$ is the eigenstate of the center of mass, and $\ket{\nu}$ the eigenstate of the relative coordinate. Since the center of mass is not affected by the interaction, the eigenstates of $H_{CM}$ are simply the standard harmonic oscillator eigenfunctions $\bra{X}\ket{n}=\frac{1}{\sqrt{2^{n}n!}}\left(\frac{m\omega}{\pi\hbar}\right)^{\frac{1}{4}}e^{-\frac{X^{2}}{2a^{2}}}H_{n}(\frac{X}{a})$, where $a=\sqrt{\frac{\hbar}{m\omega}}$ and $H_{n}$ are the Hermite polynomials, and the eigenenergies are given by $E_{CM}^{n}=\hbar\omega(n+\frac{1}{2})$. For the relative coordinate, only the even states are affected by the interaction and are given by $\bra{x}\ket{2\nu}=N_{2\nu}e^{-\frac{x^{2}}{2a^{2}}}U\left(\frac{1}{4}-\frac{E_{r}^{2\nu}}{2\hbar\omega},\frac{1}{2},\frac{x^{2}}{a^{2}}\right)$ where $N_{2\nu}$ is a normalization factor, $E_{r}^{2\nu}$ the eigenenergy and $U$ is the Kummer function \cite{Busch98}. The eigenenergies are determined by the solutions of the transcendental equation
\begin{equation}\label{trans}
    -\Tilde{g}=2\frac{\Gamma\left(-\frac{E^{2\nu}_{r}}{2\hbar\omega}+\frac{3}{4}\right)}{\Gamma\left(-\frac{E^{2\nu}_{r}}{2\hbar\omega}+\frac{1}{4}\right)},
\end{equation}
where $\Tilde{g}=\frac{g}{\sqrt{2}\hbar\omega a}$ and $\Gamma(x)$ is the gamma function. The odd eigenstates are again just harmonic oscillator states with the eigenenergies $E^{2\nu+1}_{r}=\hbar\omega(2\nu+\frac{3}{2})$.

To understand how the interaction affects the efficiency, one can note that the eigenenergies of the relative coordinate can be effectively written as 
\begin{equation}
    E_{r}^{\nu}=\hbar\omega(\nu+1/2+\epsilon(\nu,\Tilde{g})),
\end{equation}
where $\epsilon(\nu,\Tilde{g})$ is an extra energy term due to the interaction, which depends on the quantum number $\nu$ and the rescaled interaction $\Tilde{g}$. Since the interaction only affects the even states, we have $\epsilon(2\nu+1,\Tilde{g})=0$ $\forall$ $\nu$. In the limit of repulsive infinite interactions, fermionization occurs \cite{Girardeau} and the even eigenenergies asymptotically approach the next higher lying odd eigenenergies, which leads to $\epsilon(2\nu,+\infty)=1$ $\forall$ $\nu$. It is also worth noting that the contact interaction has the strongest effect on the ground state energy $\epsilon(0,\Tilde{g})\geq\epsilon(2\nu,\Tilde{g})$ $\forall$ $(\nu,\Tilde{g})$. 

\subsection{Two indistinguishable bosons}\label{3c}

\begin{figure}

\includegraphics[width=\linewidth]{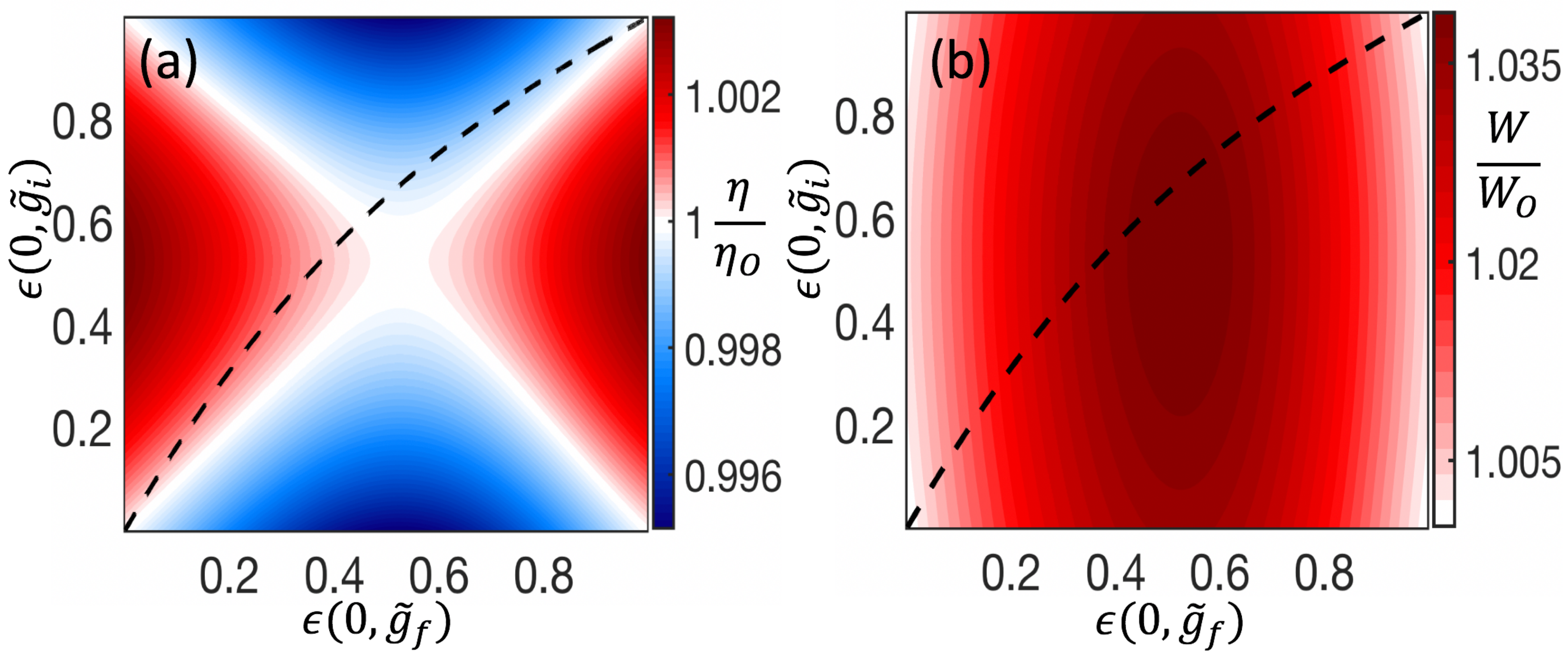}
\caption{(a) Efficiency and (b) work output normalised to their respective Otto-cycle values for an engine with a WM made of two interacting bosons as a function of $\epsilon(0,\tilde{g}_{i})$ and $\epsilon(0,\tilde{g}_{f})$. The black dash line shows the situation where the interaction is fixed $\left(g_{i}=g_{f}\right)$. Note that the efficiency converges to $\eta_{O}$ in the limit of strong interactions due to the fermionization in the system. In both plots the compression ratio is $\kappa=\frac{1}{3}$ ($\eta_{O}=\frac{2}{3}$), the cold inverse temperature is $\beta_{c}=\frac{10}{\hbar\omega_{i}}$ and the hot inverse temperature is $\beta_{h}=\frac{1}{\hbar\omega_{i}}$.}

\label{2bosons}
\end{figure}

Let us focus on the situation where the minimal QHE has a working medium consisting of two indistinguishable bosons, in which case only the states that preserve even parity can be occupied in the relative coordinate. To calculate the efficiency of the engine, we first express the heat exchanged during the hot and cold isochores as
\begin{align}
    Q_{h}=\sum_{n,\nu}E^{f}_{n,2\nu}\left(p_{n,2\nu}^{f}-p_{n,2\nu}^{i}\right),\label{heath}\\
    Q_{c}=\sum_{n,\nu}E^{i}_{n,2\nu}\left(p_{n,2\nu}^{i}-p_{n,2\nu}^{f}\right),\label{heatc}
\end{align}
where $E_{n,2\nu}^{s}=\bra{n,2\nu}H(\omega_{s},g_{s})\ket{n,2\nu}=\hbar\omega_{s}(n+2\nu+1+\epsilon(2\nu,\Tilde{g}_{s}))$ (with $s\in\{i,f\}$) and the occupation populations are given by $p_{n,2\nu}^{i}=\bra{n,2\nu}\frac{\exp\left(-\beta_{c}H(\omega_{i},g_{i})\right)}{Z(\omega_{i},g_{i},\beta_{c})}\ket{n,2\nu}$ and $p_{n,2\nu}^{f}=\bra{n,2\nu}\frac{\exp\left(-\beta_{h}H(\omega_{f},g_{f})\right)}{Z(\omega_{f},g_{f},\beta_{h})}\ket{n,2\nu}$ (where $Z(\omega,g,\beta)=\mbox{Tr}\left[\exp\left(-\beta H(\omega,g)\right)\right]$ is the partition function). From this, the efficiency can be expressed as
\begin{equation}\label{eff}
    \eta=1+\frac{Q_{c}}{Q_{h}}=1-\frac{\sum_{n,\nu}\lambda_{n,2\nu}E_{n,2\nu}^{f}\left(p_{n,2\nu}^{f}-p_{n,2\nu}^{i}\right)}{\sum_{n,\nu}E_{n,2\nu}^{f}\left(p_{n,2\nu}^{f}-p_{n,2\nu}^{i}\right)},
\end{equation}
where we have introduced $\lambda_{n,2\nu}$ as the ratio between the eigenenergies before and after the compression  \cite{Chen}
\begin{equation}\label{ratio}
    \lambda_{n,2\nu}=\frac{E_{n,2\nu}^{i}}{E_{n,2\nu}^{f}}=\kappa\frac{n+2\nu+1+\epsilon(2\nu,\Tilde{g}_{i})}{n+2\nu+1+\epsilon(2\nu,\Tilde{g}_{f})},
\end{equation}
with $\kappa=\frac{\omega_{i}}{\omega_{f}}$ being the compression ratio. From Eq.~\eqref{eff} one can see that the efficiency is influenced by the interaction through the ratio $\lambda_{n,2\nu}$, and the change of  population occupation $p_{n,2\nu}^{f}-p_{n,2\nu}^{i}$. Let us recall that the eigenstates of the harmonic oscillator for two different frequencies $\omega_{i}$ and $\omega_{f}$ are related by the scaling transformation $\bra{x}\ket{n(\omega_{f})}=\kappa^{-\frac{1}{4}}\bra{x\kappa^{-\frac{1}{2}}}\ket{n(\omega_{i})}$. Also the contact interaction described by a delta function obeys the scaling law $g\delta(\lambda x)=\frac{g}{\lambda}\delta(x)$. As a consequence, if one chooses the final interaction to be $g_{f}=g_{i}\kappa^{-\frac{1}{2}}$ (and so $\tilde{g}_{f}=\tilde{g}_{i}$), then the system will remain scale-invariant i.e. it is self-similar in space and all the eigenenergies change by the same ratio given by $\kappa$, i.~e.
\begin{equation}
    E^{i}_{n,2\nu}= \kappa E^{f}_{n,2\nu} \; \forall (n,\nu). 
\end{equation} 
In that case the efficiency is given by the Otto efficiency $\eta_{O}=1-\kappa$, which also corresponds to the efficiency of the Otto cycle of non-interacting particles. This is illustrated in Fig.~\ref{2bosons}(a), where the ratio between $\eta$ and $\eta_{O}$ is plotted as a function of $\epsilon(0,\tilde{g}_{i})$ and $\epsilon(0,\tilde{g}_{f})$ and the diagonal corresponds to the case where the WM is scale-invariant. To obtain an efficiency that differs from $\eta_{O}$, one therefore needs to consider systems where the eigenenergies do not change uniformly during the adiabatic strokes \cite{Oliveira}. To do that, we tune the initial and final interactions such that $g_{f}\neq g_{i}\kappa^{-\frac{1}{2}}$, which allows one to distinguish two possible cases. 

The first case is when the interaction weakens during the compression stroke ($\tilde{g}_{i}>\tilde{g}_{f}$), which leads to $\epsilon(2\nu,\Tilde{g}_{f})<\epsilon(2\nu,\Tilde{g_{i}})$ and $\lambda_{n,2\nu}>\kappa$ (the region above the diagonal in Fig.~\ref{2bosons}(a)). One then needs to be careful with the sign of the change of the occupation population $p_{n,2\nu}^{f}-p_{n,2\nu}^{i}$. For the excited states, the sign will be positive since at higher temperatures, the occupation population in the excited states increases. However the change of population for the ground state will be negative since it decreases for higher temperatures. Thus, depending on which terms have the largest contribution, the efficiency can be higher or lower than $\eta_{O}$. When the interactions affect the ground state more than the excited states, the change of the occupation population of the ground state is thus more important and we get $\eta>\eta_{O}$ (red area above the diagonal in Fig.~\ref{2bosons}(a)). However, when the interactions are such that the extra energy $\epsilon(2\nu,\tilde{g})$ affects significantly the excited states, the change of the occupation population for the excited states can be large enough that $\eta<\eta_{O}$ (blue area above the diagonal in Fig.~\ref{2bosons}(a)). We also indicate the situation where the interaction is fixed $g_{i}=g_{f}$ (black dashed line in Fig.~\ref{2bosons}(a)), which is similar to the situation studied in \cite{Chen}. In their case the WM is an interacting gas trapped in a box and the efficiency only decreases in the presence of the interaction. For the harmonic oscillator, however, we  observe that the interaction can enhance or hinder the performance of the engine.

The second case is when the interaction strength increases during the compression stroke ($\tilde{g}_{f}>\tilde{g}_{i}$). This implies that $\epsilon(2\nu,\tilde{g}_{f})>\epsilon(2\nu,\tilde{g}_{i})$ and $\lambda_{n,2\nu}<\kappa$. By doing the same analysis as above, we reach the opposite conclusion to the first case: if the interactions affect the ground state more we get $\eta<\eta_{O}$ (blue area below the diagonal in Fig.~\ref{2bosons}(a)), and if the interactions affect the excited states sufficiently we find $\eta>\eta_{O}$ (red area below the diagonal in Fig.~\ref{2bosons}(a)). We also highlight the area near the anti-diagonal in Fig.~\ref{2bosons}(a), where the efficiency is equal to $\eta_{O}$. This area is not exactly the anti-diagonal and corresponds to a crossover between the regime where $\eta>\eta_{O}$ and $\eta<\eta_{O}$ in which the contribution from the ground state and the excited states are such that they compensate each over and one recovers the Otto efficiency.

The maximum efficiency shown in Fig.~\ref{2bosons}(a) is $\eta\approx 1.003\eta_{0}$ and achieved for $\Tilde{g}_{i}=1.6$ and $\Tilde{g}_{f}=50$ ($\epsilon(0,\tilde{g}_{i})\approx 0.52$ and $\epsilon(0,\tilde{g}_{f})\approx 0.98$). One can see from the work output shown in panel (b) that the engine outperforms the Otto cycle of two non-interacting bosons when the final interaction takes intermediate values, while the initial interaction does not seem to influence the work output significantly. This plot also shows that the work output always exceeds $W_{O}$ and becomes equal to it when the initial and final interaction are zero or go to infinity (the four corners in Fig.~\ref{2bosons}(b)). In the infinite interaction regime this is due to the system behaving like two non-interacting fermions. The maximum work output is $W\approx1.039W_{O}$ for $\Tilde{g}_{i}=\Tilde{g}_{f}=1.6$. %It is also worth noting that if the WM is in the infinite interacting regime ($\Tilde{g}_{i}=\Tilde{g}_{f}=+\infty$), the efficiency and the work converge towards $\eta=\eta_{O}$ and $W=W_{O}$ since the system behaves like two non-interacting fermions.

\subsection{Two distinguishable particles}\label{3d}
While driving the interaction during the cycle can clearly modify the performance of the engine, the changes observed above for a working medium made from two identical bosons are not very significant and the performance of the engine stays relatively close to its non-interacting counterpart. Let us therefore next consider the situation where the working medium consists of two distinguishable  particles, for which two major differences come into play: first, as shown above, the degeneracy of the states is different in the non-interacting limit, and, second, for such a system the odd states of the energy spectrum of the relative coordinate need to be taken into account.

\begin{figure}

\includegraphics[width=\linewidth]{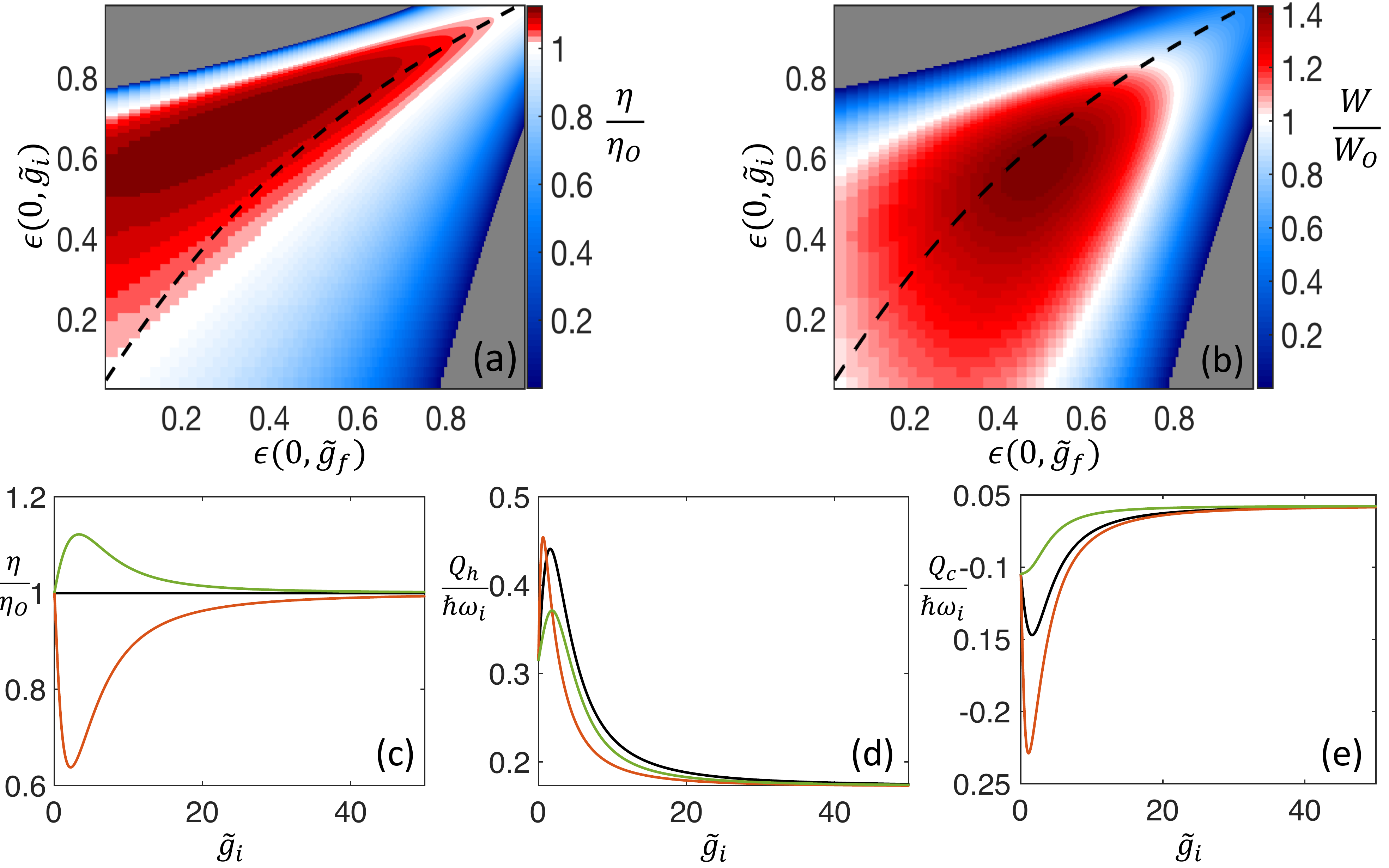}
\caption{(a) Efficiency and (b) work output normalised to
their respective Otto-cycle values for an engine with a WM
made of two interacting distinguishable particles as a function of $\epsilon(0,\tilde{g}_{i})$ and $\epsilon(0,\tilde{g}_{f})$. The gray areas correspond to interaction regimes where the system does not work as a heat engine but rather like a dissipator with $W>0$. The black dashed line shows the case where the interaction is fixed ($g_{i}=g_{f}$). (c) Efficiency normalized to the Otto efficiency and heat exchanged with the (d) hot bath and (e) cold bath as a function of the initial interaction $\Tilde{g}_{i}$, with the final interaction given by $\Tilde{g}_{f}=\Tilde{g}_{i}$ (black line), $\Tilde{g}_{f}=\frac{\Tilde{g}_{i}}{3}$ (green line) and $\Tilde{g}_{f}=3\Tilde{g}_{i}$ (orange line). The compression ratio for all plots is $\kappa=\frac{1}{3}$ ($\eta_{O}=\frac{2}{3}$), the cold inverse temperature is $\beta_{c}=\frac{10}{\hbar\omega_{i}}$ and the hot inverse temperature is $\beta_{h}=\frac{1}{\hbar\omega_{i}}$.}
\label{2d}
\end{figure}

The efficiency and work output for this engine are shown as a function of the interaction energies in Figs.~\ref{2d}(a)-(b). Like in case for indistinguishable particles, the diagonal represents the scale-invariant cycle and therefore retains the Otto efficiency, while this can be exceeded for $\epsilon(0,\tilde{g}_{i})>\epsilon(0,\tilde{g}_{f})$. Indeed, we note that interactions can noticeably improve the performance of the distinguishable cycle (note the difference in the colour scale) with the maximum efficiency $\eta\approx1.124\eta_{O}\approx0.75$ for $\Tilde{g}_{i}=3$ and $\Tilde{g}_{f}=0.8$ ($\epsilon(0,\tilde{g}_{i})\approx0.69$ and $\epsilon(0,\tilde{g}_{f})\approx0.34$), and maximum work $W\approx1.43W_{O}$ for $\Tilde{g}_{i}=1.95$ and $\Tilde{g}_{f}=1.4$ ($\epsilon(0,\tilde{g}_{i})\approx0.58$ and $\epsilon(0,\tilde{g}_{f})\approx0.48$). In contrast to indistinguishable bosons the efficiency is always reduced when the final interaction is larger than the initial one, $\tilde{g}_f>\tilde{g}_i$. Furthermore, the work output can be significantly lower than for a cycle with a non-interacting particles, and we note that the WM can act as a dissipator ($W>0$) for combinations of strong and weak interactions, $(\tilde{g}_i\approx 0,\tilde{g}_f\rightarrow \infty)$ and $(\tilde{g}_i\rightarrow \infty,\tilde{g}_f\approx 0)$, indicated by the grey regions in Figs.\ref{2d}(a)-(b). We have also calculated the efficiency for higher temperatures and have observed the same general behavior, however, the variations of the efficiency and the work output become less pronounced.

To illustrate and better understand the behaviour of the QHE with distinguishable particles, we calculate the efficiency as a function of the initial interaction $\tilde{g}_{i}$ while tuning the final interaction such that $\tilde{g}_{f}=\alpha\tilde{g}_{i}$, with $\alpha$ fixed. Fig.~\ref{2d}(c) shows the efficiency for three different values of $\alpha$ ($\alpha=1$, $3$ and $\frac{1}{3}$) and one can clearly see that the changes in the efficiency are more significant and also very different from the setting using indistinguishable particles. The efficiency can be enhanced when $\Tilde{g}_{f}<\Tilde{g}_{i}$, while it is significantly reduced when $\Tilde{g}_{f}>\Tilde{g}_{i}$. To understand this one can consider the amount of heat exchanged with the hot and cold baths as shown in Fig.~\ref{2d}(d,e). In Fig.~\ref{2d}(d) we see that the presence of the interaction increases the amount of heat received by the hot bath in all three cases,  for weak and intermediate values of $\Tilde{g}_{i}$. However it decreases for large $\Tilde{g}_i$ and reaches a limit that is approximately half that 
%close to half of the amount of heat received by
of the non-interacting working medium. In this limit the even eigenstates in the relative coordinate approach the next higher-lying odd eigenstates and thus the spectrum becomes doubly degenerated for distinguishable particles which implies less heat is required for the WM to thermalize. 

We observe similar behavior for $Q_{c}$ in the large $\tilde{g}_i$ limit (Fig.~\ref{2d}(e)), however for weak and intermediate values of $\Tilde{g}_{i}$ the amount of heat dissipated in the cold bath is significantly larger when $\tilde{g}_{f}=3\tilde{g}_{i}$, %(more than twice at the extremum) 
which is the reason for the decreasing efficiency. For the same initial interaction $\Tilde{g}_{i}$ the best strategy to reduce energy loss in the cold bath is therefore to choose a weaker final interaction $\Tilde{g}_{f}$. This allows the statistics at the inverse temperature $\beta_{h}$ to be closer to the initial statistics of the WM and thus the change of the occupation population $p_{n,\nu}^{i}-p_{n,\nu}^{f}$ becomes smaller in magnitude such that less heat is released during the cold isochore. The changes in the performance of the QHE are more extreme when the baths are at low temperatures because the WM is more affected by finite interactions. 

\begin{figure}
\includegraphics[width=\linewidth]{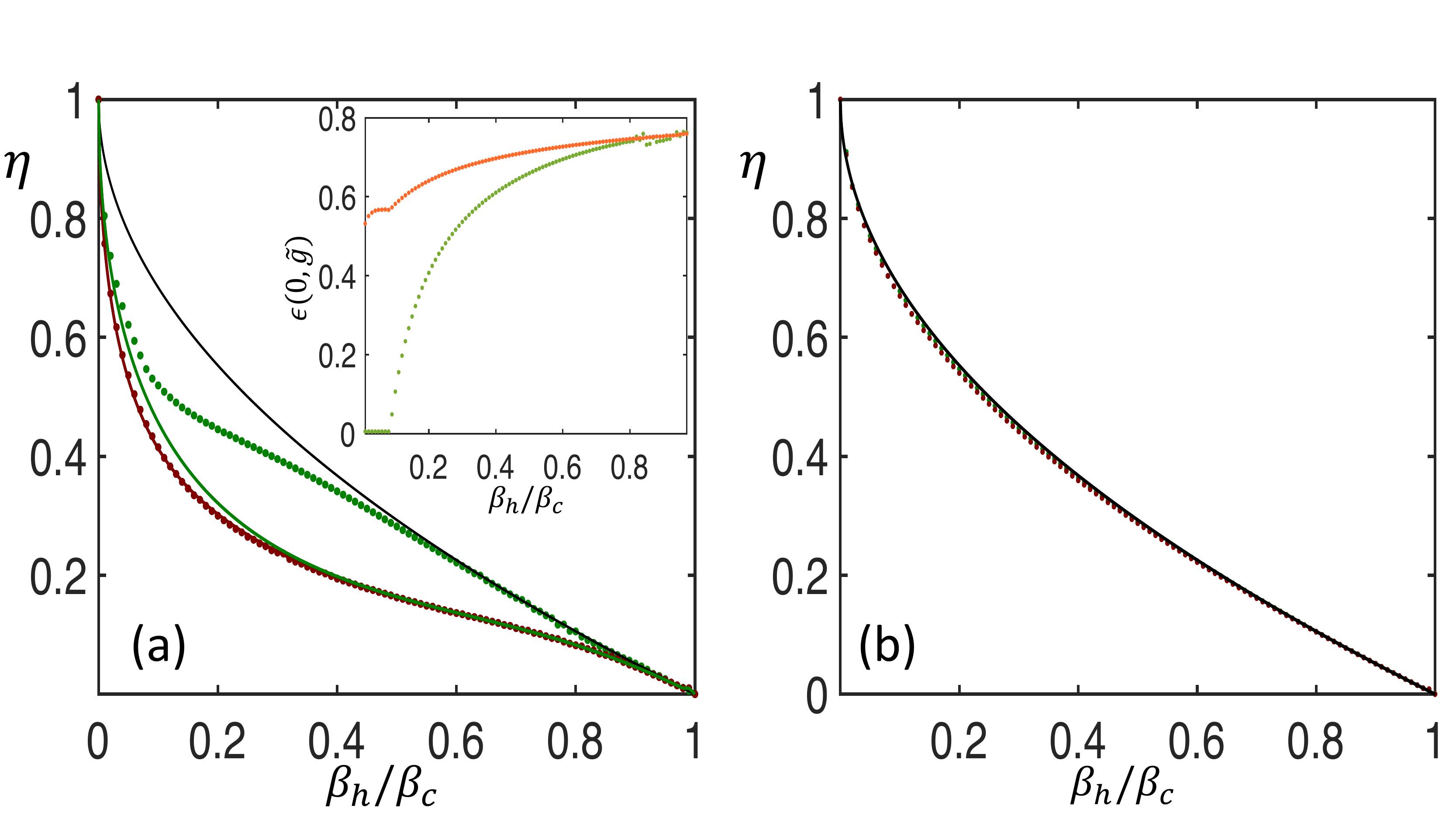}
\caption{Efficiency at maximum work output (EMW) for the engine using two interacting bosons (brown dots)  and two interacting distinguishable particles (green dots) for two different temperature regimes. Panel (a) shows the efficiency calculated in the low temperature regime with $\beta_{c}\hbar\omega_{i}=10$  and panel (b) in the intermediate temperature regime with $\beta_{c}\hbar\omega_{i}=1$. In both panels the back line corresponds to the Curzon-Ahlborn bound $\eta_{CA}$.
The inset in panel (a) shows the corresponding extra energies $\epsilon(0,\tilde{g}_{i})$ (orange dots) and $\epsilon(0,\tilde{g}_{f})$ (green dots) for the case of two distinguishable particles. The values of $\epsilon(0,g_{f})$ we obtained for $\beta_{h}/\beta_{c}\gtrsim0.8$ become less accurate because the work output starts to vanish in this regime. We also show the efficiency at maximum work output of two non-interacting distinguishable particles (green line) and two non-interacting bosons (brown line) at the low temperature regime (a). Note the EMW for two interacting bosons is extremely close to that of the two non-interacting bosons.  
}
\label{etamax}
\end{figure}

It is also insightful to look at the efficiency at maximum work output (EMW) which allows to quantify the engine performance for different temperature scales and helps understanding finite time dynamics. We compare the EMW to the Curzon-Ahlborn bound \cite{CA,Leff}
\begin{equation}
    \eta_{CA}=1-\sqrt{\frac{\beta_{h}}{\beta_{c}}}\,,
\end{equation}
which originally corresponded to the efficiency at maximum power of the endoreversible Carnot cycle. However it has been shown that it also holds for the quantum Otto cycle at high temperature and in the adiabatic limit \cite{Rezek}. We calculate the efficiency by maximizing the work output over $\kappa$, $\Tilde{g}_{i}$ and $\Tilde{g_{f}}$ for two fixed cold bath temperatures $\beta_{c}\hbar\omega_{i}=1$ and $\beta_{c}\hbar\omega_{i}=10$ (see Fig.~\ref{etamax}). We also compare the EMW of both engines, with distinguishable and indistinguishable interacting working media, with their non-interacting counterparts in the low temperature regime $\beta_{c}\hbar\omega_{i}=10$ in Fig.~\ref{etamax}(a). One can see that for the non-interacting engines, this quantity is far below the CA bound, which is due to the fact that at low temperature the WM dissipates a large amount of energy into the cold bath in order to close the cycle. At the same time the efficiency for non-interacting bosons is the lowest  as its statistics makes low energy states more favorable than in the distinguishable case. The EMW of the interacting bosons is extremely close to that of non-interacting bosons, which is not surprising since the influence of the interaction on the engine performance is very small (as seen in Fig.~\ref{2bosons}). However for the two distinguishable particles, it is vastly improved by the presence of the interaction and even coincides with the CA bound for $\beta_{h}/\beta_{c}\gtrsim0.5$. One can see that the gap between the efficiency of bosons and distinguishable particles decreases when the temperature of the hot bath is large, $\beta_h/\beta_c\rightarrow 0$, as their statistics become identical and are given by the classical Maxwell-Boltzmann distribution. Also the energy scales of the hot bath dwarf that of the cold bath and therefore the influence of the initial interaction $\tilde{g}_{i}$ becomes negligible. 

This is highlighted in the inset of Fig.~\ref{etamax}(a) where we show the corresponding extra energy $\epsilon(0,\tilde{g}_{i})$ and $\epsilon(0,\tilde{g}_{f})$ for the case of two distinguishable particles. As expected from our previous analysis, the final interaction $\tilde{g}_{f}$ has to be smaller than $\tilde{g}_{i}$ in order to improve the performance of the engine. The optimal final interaction $\tilde{g}_{f}$ decreases when the temperature of the hot bath increases and becomes zero for $\beta_{h}/\beta_{c}\lesssim0.08$. As we have already mentioned, the work output in the high temperature limit is mostly dictated by the energy of the WM at the inverse temperature $\beta_{h}$. Therefore the influence of the interaction becomes negligible since the particles behave like non-interacting classical particles with an energy approximately given by $\beta_{h}^{-1}$. Moreover, from the preceding analysis, we know that the work output is significantly improved for non-zero and finite interaction strengths (Fig.~\ref{2d}(b)) and we can thus conclude that the interactions start to influence the performance of the engine for $\beta_{h}/\beta_{c}\gtrsim0.08$. We can also observe this in the efficiencies of the different cycles which start to deviate from each other at around this temperature (see Fig.~\ref{etamax}(a)). %zero for high temperatures and only becomes non-zero for $\frac{\beta_{h}}{\beta_{c}}\gtrsim 0.08$, corresponding to the temperature regime where the final interaction influences the performance of the engine. 

When $\beta_{h}/\beta_{c}\rightarrow 1$ both $\tilde{g}_{i}$ and $\tilde{g}_{f}$ tend to the same limit in which the cycle is scale invariant. We note the slight irregularities can be seen for the behavior of $\epsilon(0,g_{f})$ in the inset of Fig.~\ref{etamax}(a) when $\beta_{h}/\beta_{c}\gtrsim0.8$. This is due to numerical issues, as the optimization algorithm has difficulties in finding the maximum work output when the temperatures of both baths are close, and therefore the work output starts to vanish. Regardless, in this regime the efficiencies of each cycle converge to the Curzon-Ahlborn bound as expected. Finally, we show that for a larger temperature of the cold bath $\beta_{c}\hbar\omega_{i}=1$, the EMW for both working media are exactly equal to $\eta_{CA}$ (see Fig.~\ref{etamax}(b)). In this temperature regime, the effect of the short-range interaction becomes negligible and the particles behave like a non-interacting ideal and classical gas. 

\subsection{Three particles working medium}\label{3e}
\begin{figure}

\includegraphics[width=\linewidth]{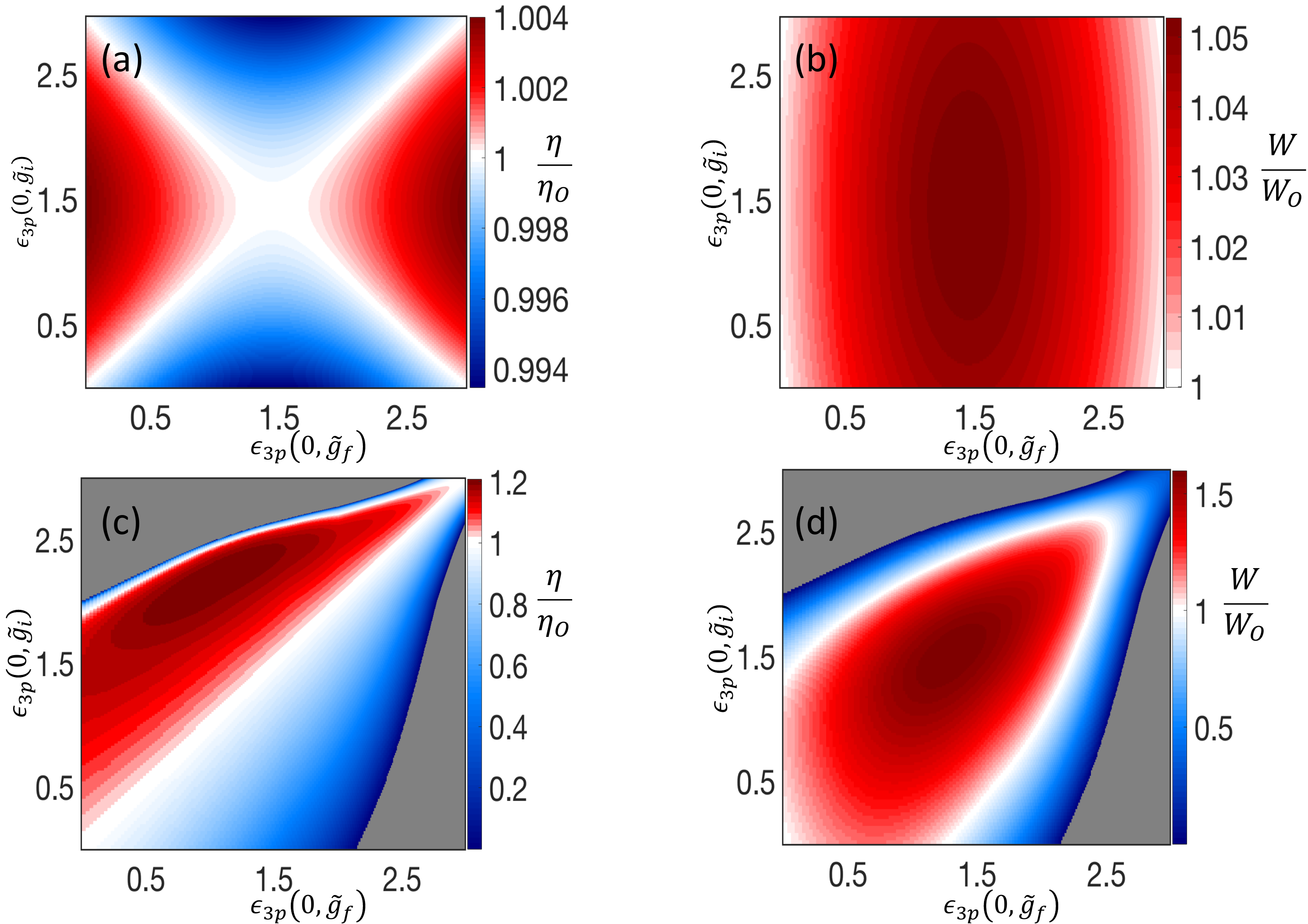}
\caption{Efficiency and work output normalised to their respective Otto values as a function of $\epsilon_{3P}(0,\tilde{g}_{i})$ and $\epsilon_{3P}(0,\tilde{g}_{f})$ for a WM consisting of (a)-(b) three indistinguishable bosons and (c)-(d) three distinguishable particles. 
%(c) Efficiency normalised to the Otto efficiency as a function of $\epsilon_{3P}(0,\tilde{g}_{i})$ and $\epsilon_{3P}(0,\tilde{g}_{f})$ for three distinguishable particles. (d) Corresponding work output normalized to the work output of its non-interacting counterpart engine.     
}
\label{3p}
\end{figure}

In order to show how the influence of the interaction on the engine performance scales with the number of particles, we next extend the analysis to a three particle system. In that case the Hamiltonian can no longer be analytically solved and a numerical method such as exact diagonalization is required to calculate the quantities of interest. To compare with the two particle engines we consider equivalent temperatures and compression ratios, and also define the interaction energy $\epsilon_{3P}(n,\tilde{g})$ in a similar way. With $n$ being the quantum index that characterizes the $n$-th three particle eigenstate, the ground state interaction energy term is given by
\begin{equation}
    \epsilon_{3P}(0,\tilde{g})=\frac{\bra{0}H(\omega,g)\ket{0}}{\hbar\omega}-\frac{3}{2}.
\end{equation}
This excess energy is such that $\epsilon_{3P}(0,0)=0$ and $\epsilon_{3P}(0,+\infty)=3$ and again allows us to quantify the interaction strength in the system. In Fig.~\ref{3p} we show the efficiency and work output for both indistinguishable and distinguishable particles. We note that the optimal interactions needed for maximizing performance does not significantly change compared to the two particle case, however, the degree of enhancement is marginally increased, with a gain of $0.1\%$ for the maximum efficiency ($\eta\approx1.004\eta_{O}\approx0.668$ for $\epsilon_{3P}(0,\Tilde{g}_{i})\approx1.64$ and $\epsilon_{3P}(0,\tilde{g}_{f})\approx2.96$) and $1.1\%$ for the work output ($W\approx1.05W_{O}$ for $\epsilon_{3P}(0,\tilde{g}_{i})=\epsilon_{3P}(0,\tilde{g}_{f})\approx1.64$). However, the engine using a WM of distinguishable particles shows a more significant enhancement of the performance as the number of states that are not affected by the interaction is much larger than in the two particle case (see Fig.~\ref{d_vs_b}(a)-(b)) allowing for a more efficient work extraction process. The resulting maximum efficiency and work output, $\eta\approx1.21\eta_{O}\approx0.807$ for $\epsilon_{3P}(0,\tilde{g}_{i})\approx2.48$ and $\epsilon_{3P}(0,\tilde{g}_{f}=0.8)\approx1.49$ and $W\approx1.59W_{O}$ for $\epsilon_{3P}(0,\tilde{g}_{i})\approx1.73$ and $\epsilon_{3P}(0,\tilde{g}_{f})\approx1.49$, allowing for gains of $8.6\%$ and $16\%$ respectively over the distinguishable two particle engine. This highlights the important role the density of states plays in the performance and how this can be modified by the statistics.

\begin{figure}
\includegraphics[width=0.9\linewidth]{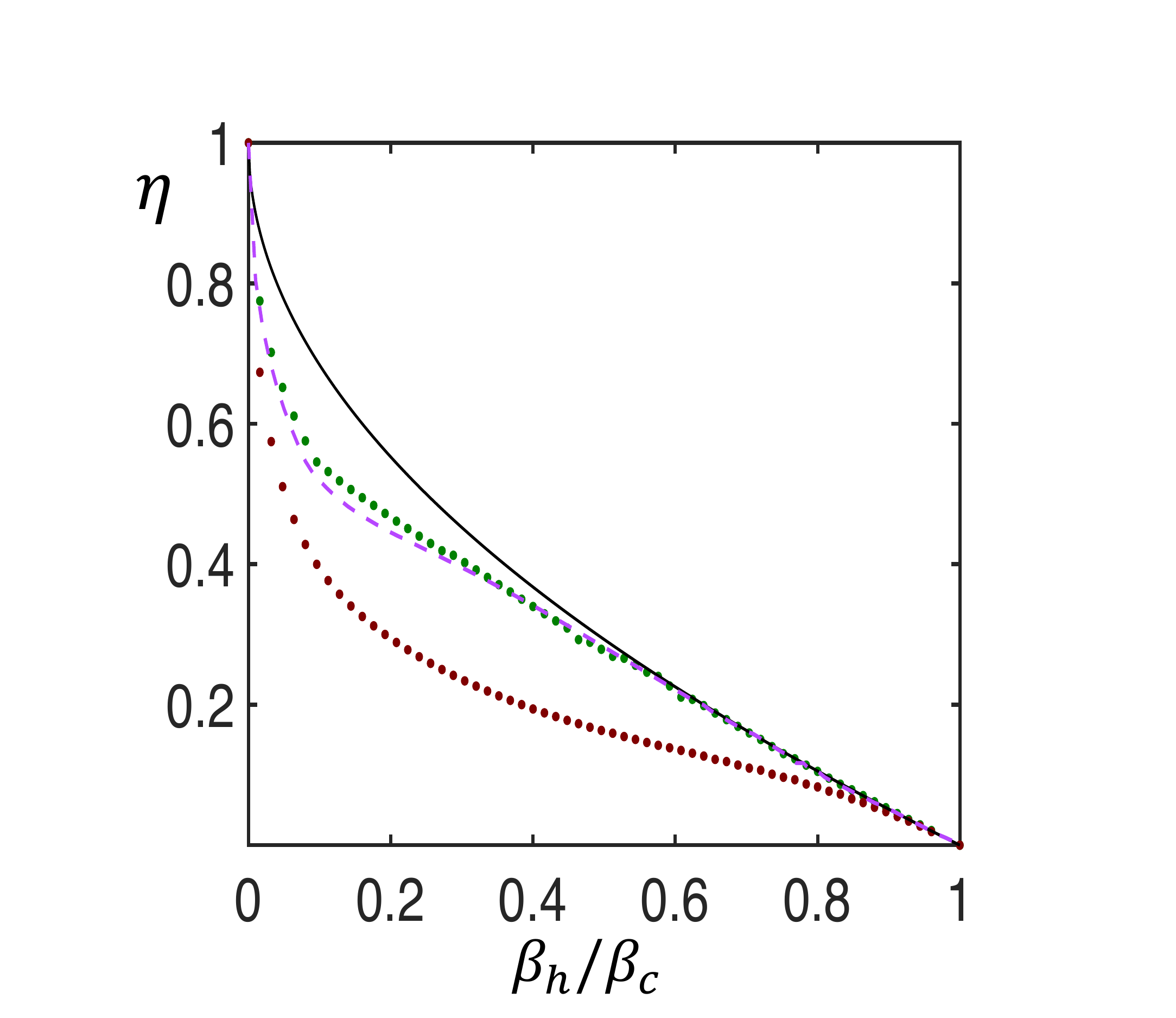}
\caption{Efficiency at maximum work output for three interacting bosons (brown dots) and three distinguishable particles (green dots) in the low temperature regime $\beta_{c}\hbar\omega_{i}=10$. The black line corresponds to the Curzon-Ahlborn bound $\eta_{CA}$ and the purple dashed line shows the two distinguishable interacting particles case.
}
\label{etamax3p}
\end{figure}

Finally, we show in Fig.~\ref{etamax3p} the EMW and compare the cases for two and three distinguishable particles. While the EMWs for the engines with indistinguishable particles are very similar, one can note a slight enhancement of the efficiency for the WM made from indistinguishable particles in the intermediate temperature regime.

\section{Finite time dynamics}\label{4}
So far we studied the performance of the engine in the adiabatic limit, which results in a vanishing power output due to the long time-scale of the strokes. While reducing the duration of the strokes increases the power, it will, however, inevitably lead to the decrease of the efficiency due to the generation of irreversibility and inner friction \cite{Rezek,Alecce_2015,Feldmann,Feldmann_2003,Cakmak_2017,Esposito,Andresen,Whitney,Shiraishi,Lin,Abah_2016}. In order to understand the trade-off between the power and the efficiency of the engine, we next study the compression and expansion strokes at finite-time for an engine made from two distinguishable particles. For this we consider three cases. First, we change the interaction between two values that are known to give a boost to the efficiency and work output in the adiabatic limit (the optimal case), and second we drive the interaction in such a way that the WM remains scale-invariant, for which the efficiency in the adiabatic limit is given by $\eta_{O}$. As a third case we consider two non-interacting particles in order to benchmark our results. As our focus is on the non-adiabatic excitation during the compression and expansion strokes, we do not consider the dynamics of the isochoric strokes. The duration $\tau$ of the compression and expansion stokes are taken to be same and  for the optimal case the ramps for the time-dependent protocols for the trap and interaction strengths are given by
\begin{equation}
    f(t)=f(0)+10\Delta f\left(\frac{t}{\tau}\right)^{3}-15\Delta f\left(\frac{t}{\tau}\right)^{4}+6\Delta f\left(\frac{t}{\tau}\right)^{5},
    \label{ramp}
\end{equation}
where $\Delta f=f(\tau)-f(0)$ for $f=\{g,\omega\}$. In the case of scale-invariant dynamics the interaction strength is connected to the trap frequency through $g(t)=g(0)\sqrt{\frac{\omega(t)}{\omega(0)}}$ and we choose $\omega(t)$ to be given by Eq.~\eqref{ramp}. To quantify the performance of the engine at finite-time, we calculate the efficiency $\eta$ and also the effective power (EP) of the engine defined as
\begin{equation}
   P(\tau)=-\frac{W(\tau)}{2\tau}. 
\end{equation}
While, the latter does not strictly correspond to the power of the engine since we only consider the total duration of the compression and expansion strokes $2\tau$, we always assume that the WM fully thermalizes during the isochoric stokes for a short fixed time. The EP then tells us how the power of the engine is affected by non-adiabatic excitations created during the work strokes, and if the duration of the isochoric stokes is short enough, it corresponds to the first approximation of the engine power. Finally we also quantify the irreversibility of the cycle by calculating the irreversible work

\begin{equation}
    W_{irr}(\tau)=W(\tau)-W_{ad},
\end{equation}
where $W_{ad}$ is the work output of the engine in the adiabatic limit.

\begin{figure}
\includegraphics[width=\linewidth]{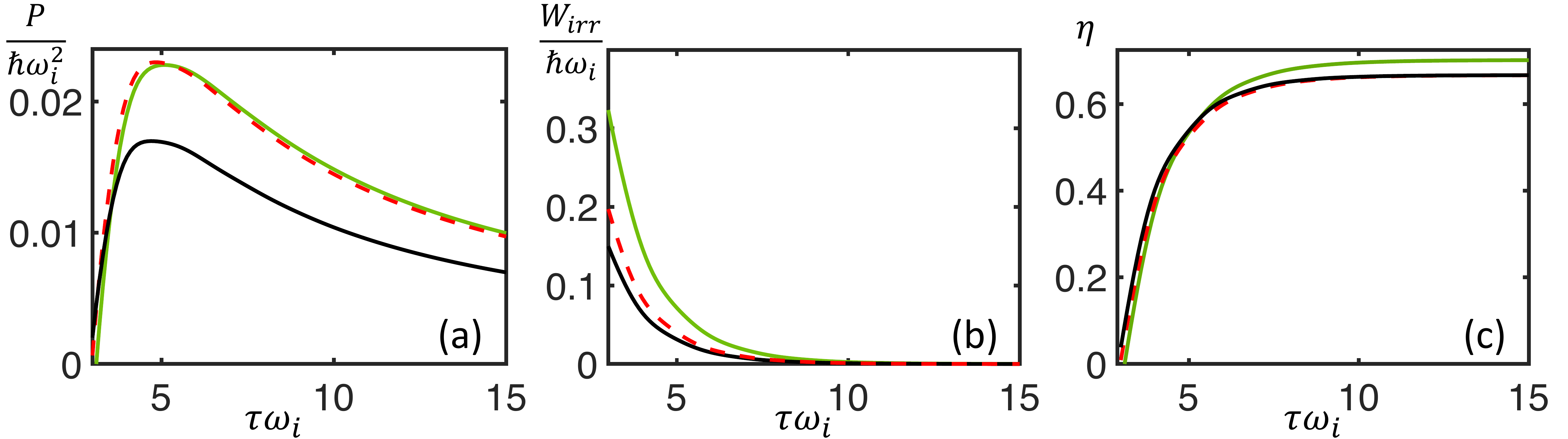}
\caption{(a) Effective power output (EP), (b) irreversible work and (c) efficiency  as a function of $\tau$. The green line corresponds to the optimal case, the red dashed line corresponds to the scale-invariant case and the black line to the non-interacting case.  The compression ratio is $\kappa=\frac{1}{3}$, the cold inverse temperature is $\beta_{c}=\frac{10}{\hbar\omega_{i}}$ and the hot inverse temperature is $\beta_{h}=\frac{1}{\hbar\omega_{f}}$. For the optimal case the interactions are $\tilde{g}_{i}=1.95$ and $\tilde{g}_{f}=1.4$ such that the efficiency in the adiabatic limit is $\eta\approx0.7$, and for the scale-invariant case the interactions are $\tilde{g}_{i}=\tilde{g}_{f}=1.95$. 
}
\label{finite_time2p}
\end{figure}

The EP, the irreversible work and the efficiency as functions of $\tau$ are shown in Fig.~\ref{finite_time2p}. Compared to the non-interacting case, using an interacting system in a cycle of finite duration provides a significant boost to the power in both the optimal case and the scale-invariant case. Moreover, one can see that the EP in the optimal case is larger than in the scale-invariant case for longer stroke durations, which is due to the work output being larger in the adiabatic limit. However for fast strokes the EP for the scale-invariant case becomes larger and the maximum value is reached at a shorter time than in the optimal case. While this is a small effect, it is consistent with the greater amount of irreversible work being generated in the optimal case (see Fig.~\ref{finite_time2p}(b)). It can also be seen from the fact  that even if the efficiency in the adiabatic limit in the optimal case is greater, it decreases faster for shorter times than in the scale-invariant case (see Fig.~\ref{finite_time2p}(c)). Finally, one can note that the irreversible work generated by driving the interaction in the scale-invariant case is not very significant and therefore the efficiency in this case stays relatively close to the efficiency of the non-interacting case, even for short times.

\section{Conclusion}\label{5}
In summary, we have investigated a quantum Otto heat engine, where the working medium is an interacting quantum system and the interaction is driven at the same time as the trap frequency of the system during the adiabatic strokes. In our case the interaction is a short-ranged and repulsive, and we have shown that it allows to change the performance of the engine and outperform a quantum Otto heat engine realized with non-interacting particles. 

The performance of the engine is modified for two reasons: the first is that the interaction does not have the same effect on all eigenstates and therefore the eigenenergies do not change uniformly during the adiabatic strokes. This is in contrast to models where the energy shift due to the interactions is the same for all states, such as the Calogero-Sutherland model, and which do not diverge from the Otto efficiency in the adiabatic limit \cite{Myers, Jaramillo_2016}. This means that our engine shows a different behavior for finite interaction values. The second reason is that the interaction affects the energy distribution which allows, for example, to lose less energy during the cold isochore. The interaction has however only a very small effect in the case of indistinguishable bosons, while it can significantly modify the engine performance for distinguishable particles due to the presence of odd eigenstates states that are not affected by the interaction. While the tiny influence of the interaction in the indistinguishable case would be extremely difficult to experimentally observe, the significant improvement obtained in the distinguishable case can be expected to be experimentally measurable. Furthermore, we have found that the interactions mostly matter at low temperatures, and that increasing the number of particles does not seem to modify the influence of the interaction on the engine performance. In fact, it continues to enhance the work output and the efficiency in the case of distinguishable particles. 

We have also studied non-adiabatic engine cycles of finite duration and quantified the trade-off between the power and efficiency. We have shown that driving the trap and the interaction by keeping the WM scale-invariant generates less irreversible work which allows better performances in terms of power but also efficiency at short time scales. More generally, using an interacting system in a finite time cycle is significantly advantageous in terms of power. Our work therefore shows the potential for developing QHEs that possess multiple control parameters which can be changed during the work strokes.

A number of interesting questions immediately emerge from our work. While we have only considered repulsive interactions, attractive interactions could potentially also lead to higher efficiencies. Furthermore, since the external potential has a strong influence on the spectrum, optimising the trapping potential would be a insightful study relating to the quantum nature of the engine \cite{Chen}.  Finally, considering long range interactions within the working medium \cite{Beau_2020} could lead to engines that show different behaviour at higher temperatures.

\section*{Acknowledgements}
This work was supported by the Okinawa Institute of Science and Technology Graduate University, and used the computing resources of the Scientific Computing and Data Analysis section at OIST. TF acknowledges support from JSPS under the grant KAKENHI-21K13856. TF and TB are also supported by JST Grant Number JPMJPF2221.

%%%%%%%%%%%%%%
% References %
%%%%%%%%%%%%%%

%\bibliographystyle{physics-bibstyle}
%\bibliographystyle{apsrev4-1}
\bibliography{main}

\begin{thebibliography}{51}%
\makeatletter
\providecommand \@ifxundefined [1]{%
 \@ifx{#1\undefined}
}%
\providecommand \@ifnum [1]{%
 \ifnum #1\expandafter \@firstoftwo
 \else \expandafter \@secondoftwo
 \fi
}%
\providecommand \@ifx [1]{%
 \ifx #1\expandafter \@firstoftwo
 \else \expandafter \@secondoftwo
 \fi
}%
\providecommand \natexlab [1]{#1}%
\providecommand \enquote  [1]{``#1''}%
\providecommand \bibnamefont  [1]{#1}%
\providecommand \bibfnamefont [1]{#1}%
\providecommand \citenamefont [1]{#1}%
\providecommand \href@noop [0]{\@secondoftwo}%
\providecommand \href [0]{\begingroup \@sanitize@url \@href}%
\providecommand \@href[1]{\@@startlink{#1}\@@href}%
\providecommand \@@href[1]{\endgroup#1\@@endlink}%
\providecommand \@sanitize@url [0]{\catcode `\\12\catcode `\$12\catcode
  `\&12\catcode `\#12\catcode `\^12\catcode `\_12\catcode `\%12\relax}%
\providecommand \@@startlink[1]{}%
\providecommand \@@endlink[0]{}%
\providecommand \url  [0]{\begingroup\@sanitize@url \@url }%
\providecommand \@url [1]{\endgroup\@href {#1}{\urlprefix }}%
\providecommand \urlprefix  [0]{URL }%
\providecommand \Eprint [0]{\href }%
\providecommand \doibase [0]{https://doi.org/}%
\providecommand \selectlanguage [0]{\@gobble}%
\providecommand \bibinfo  [0]{\@secondoftwo}%
\providecommand \bibfield  [0]{\@secondoftwo}%
\providecommand \translation [1]{[#1]}%
\providecommand \BibitemOpen [0]{}%
\providecommand \bibitemStop [0]{}%
\providecommand \bibitemNoStop [0]{.\EOS\space}%
\providecommand \EOS [0]{\spacefactor3000\relax}%
\providecommand \BibitemShut  [1]{\csname bibitem#1\endcsname}%
\let\auto@bib@innerbib\@empty
%</preamble>
\bibitem [{\citenamefont {Vinjanampathy}\ and\ \citenamefont
  {Anders}(2016)}]{Sai}%
  \BibitemOpen
  \bibfield  {author} {\bibinfo {author} {\bibfnamefont {S.}~\bibnamefont
  {Vinjanampathy}}\ and\ \bibinfo {author} {\bibfnamefont {J.}~\bibnamefont
  {Anders}},\ }\bibfield  {title} {\bibinfo {title} {Quantum thermodynamics},\
  }\href {https://doi.org/10.1080/00107514.2016.1201896} {\bibfield  {journal}
  {\bibinfo  {journal} {Contemporary Physics}\ }\textbf {\bibinfo {volume}
  {57}},\ \bibinfo {pages} {545} (\bibinfo {year} {2016})}\BibitemShut
  {NoStop}%
\bibitem [{\citenamefont {Alicki}\ and\ \citenamefont
  {Kosloff}(2018)}]{Alicki}%
  \BibitemOpen
  \bibfield  {author} {\bibinfo {author} {\bibfnamefont {R.}~\bibnamefont
  {Alicki}}\ and\ \bibinfo {author} {\bibfnamefont {R.}~\bibnamefont
  {Kosloff}},\ }\bibinfo {title} {{Introduction to Quantum Thermodynamics:
  History and Prospects}},\ in\ \href
  {https://doi.org/10.1007/978-3-319-99046-0_1} {\emph {\bibinfo {booktitle}
  {Thermodynamics in the Quantum Regime: Fundamental Aspects and New
  Directions}}},\ \bibinfo {editor} {edited by\ \bibinfo {editor}
  {\bibfnamefont {F.}~\bibnamefont {Binder}}, \bibinfo {editor} {\bibfnamefont
  {L.~A.}\ \bibnamefont {Correa}}, \bibinfo {editor} {\bibfnamefont
  {C.}~\bibnamefont {Gogolin}}, \bibinfo {editor} {\bibfnamefont
  {J.}~\bibnamefont {Anders}},\ and\ \bibinfo {editor} {\bibfnamefont
  {G.}~\bibnamefont {Adesso}}}\ (\bibinfo  {publisher} {Springer International
  Publishing},\ \bibinfo {address} {Cham},\ \bibinfo {year} {2018})\ pp.\
  \bibinfo {pages} {1--33}\BibitemShut {NoStop}%
\bibitem [{\citenamefont {Kosloff}\ and\ \citenamefont
  {Levy}(2014)}]{Kosloff2014}%
  \BibitemOpen
  \bibfield  {author} {\bibinfo {author} {\bibfnamefont {R.}~\bibnamefont
  {Kosloff}}\ and\ \bibinfo {author} {\bibfnamefont {A.}~\bibnamefont {Levy}},\
  }\bibfield  {title} {\bibinfo {title} {{Quantum Heat Engines and
  Refrigerators: Continuous Devices}},\ }\href
  {https://doi.org/10.1146/annurev-physchem-040513-103724} {\bibfield
  {journal} {\bibinfo  {journal} {Annual Review of Physical Chemistry}\
  }\textbf {\bibinfo {volume} {65}},\ \bibinfo {pages} {365} (\bibinfo {year}
  {2014})}\BibitemShut {NoStop}%
\bibitem [{\citenamefont {Binder}\ \emph {et~al.}(2018)\citenamefont {Binder},
  \citenamefont {Correa}, \citenamefont {Gogolin}, \citenamefont {Anders},\
  and\ \citenamefont {Adesso}}]{Binder_2018}%
  \BibitemOpen
  \bibinfo {editor} {\bibfnamefont {F.}~\bibnamefont {Binder}}, \bibinfo
  {editor} {\bibfnamefont {L.~A.}\ \bibnamefont {Correa}}, \bibinfo {editor}
  {\bibfnamefont {C.}~\bibnamefont {Gogolin}}, \bibinfo {editor} {\bibfnamefont
  {J.}~\bibnamefont {Anders}},\ and\ \bibinfo {editor} {\bibfnamefont
  {G.}~\bibnamefont {Adesso}},\ eds.,\ \href
  {https://doi.org/https://doi.org/10.1007/978-3-319-99046-0} {\emph {\bibinfo
  {title} {Thermodynamics in the Quantum Regime: Fundamental Aspects and New
  Directions}}}\ (\bibinfo  {publisher} {Springer International Publishing},\
  \bibinfo {address} {Cham},\ \bibinfo {year} {2018})\BibitemShut {NoStop}%
\bibitem [{\citenamefont {Abah}\ \emph {et~al.}(2012)\citenamefont {Abah},
  \citenamefont {Ro\ss{}nagel}, \citenamefont {Jacob}, \citenamefont {Deffner},
  \citenamefont {Schmidt-Kaler}, \citenamefont {Singer},\ and\ \citenamefont
  {Lutz}}]{Abah2012}%
  \BibitemOpen
  \bibfield  {author} {\bibinfo {author} {\bibfnamefont {O.}~\bibnamefont
  {Abah}}, \bibinfo {author} {\bibfnamefont {J.}~\bibnamefont {Ro\ss{}nagel}},
  \bibinfo {author} {\bibfnamefont {G.}~\bibnamefont {Jacob}}, \bibinfo
  {author} {\bibfnamefont {S.}~\bibnamefont {Deffner}}, \bibinfo {author}
  {\bibfnamefont {F.}~\bibnamefont {Schmidt-Kaler}}, \bibinfo {author}
  {\bibfnamefont {K.}~\bibnamefont {Singer}},\ and\ \bibinfo {author}
  {\bibfnamefont {E.}~\bibnamefont {Lutz}},\ }\bibfield  {title} {\bibinfo
  {title} {{Single-Ion Heat Engine at Maximum Power}},\ }\href
  {https://doi.org/10.1103/PhysRevLett.109.203006} {\bibfield  {journal}
  {\bibinfo  {journal} {Phys. Rev. Lett.}\ }\textbf {\bibinfo {volume} {109}},\
  \bibinfo {pages} {203006} (\bibinfo {year} {2012})}\BibitemShut {NoStop}%
\bibitem [{\citenamefont {Scovil}\ and\ \citenamefont
  {Schulz-DuBois}(1959)}]{Scovil}%
  \BibitemOpen
  \bibfield  {author} {\bibinfo {author} {\bibfnamefont {H.~E.~D.}\
  \bibnamefont {Scovil}}\ and\ \bibinfo {author} {\bibfnamefont {E.~O.}\
  \bibnamefont {Schulz-DuBois}},\ }\bibfield  {title} {\bibinfo {title}
  {{Three-Level Masers as Heat Engines}},\ }\href
  {https://doi.org/10.1103/PhysRevLett.2.262} {\bibfield  {journal} {\bibinfo
  {journal} {Phys. Rev. Lett.}\ }\textbf {\bibinfo {volume} {2}},\ \bibinfo
  {pages} {262} (\bibinfo {year} {1959})}\BibitemShut {NoStop}%
\bibitem [{\citenamefont {Geva}\ and\ \citenamefont {Kosloff}(1992)}]{Geva}%
  \BibitemOpen
  \bibfield  {author} {\bibinfo {author} {\bibfnamefont {E.}~\bibnamefont
  {Geva}}\ and\ \bibinfo {author} {\bibfnamefont {R.}~\bibnamefont {Kosloff}},\
  }\bibfield  {title} {\bibinfo {title} {{A quantum‐mechanical heat engine
  operating in finite time. A model consisting of spin‐1/2 systems as the
  working fluid}},\ }\href {https://doi.org/10.1063/1.461951} {\bibfield
  {journal} {\bibinfo  {journal} {The Journal of Chemical Physics}\ }\textbf
  {\bibinfo {volume} {96}},\ \bibinfo {pages} {3054} (\bibinfo {year}
  {1992})}\BibitemShut {NoStop}%
\bibitem [{\citenamefont {Humphrey}\ \emph {et~al.}(2002)\citenamefont
  {Humphrey}, \citenamefont {Newbury}, \citenamefont {Taylor},\ and\
  \citenamefont {Linke}}]{Humphrey}%
  \BibitemOpen
  \bibfield  {author} {\bibinfo {author} {\bibfnamefont {T.~E.}\ \bibnamefont
  {Humphrey}}, \bibinfo {author} {\bibfnamefont {R.}~\bibnamefont {Newbury}},
  \bibinfo {author} {\bibfnamefont {R.~P.}\ \bibnamefont {Taylor}},\ and\
  \bibinfo {author} {\bibfnamefont {H.}~\bibnamefont {Linke}},\ }\bibfield
  {title} {\bibinfo {title} {{Reversible Quantum Brownian Heat Engines for
  Electrons}},\ }\href {https://doi.org/10.1103/PhysRevLett.89.116801}
  {\bibfield  {journal} {\bibinfo  {journal} {Phys. Rev. Lett.}\ }\textbf
  {\bibinfo {volume} {89}},\ \bibinfo {pages} {116801} (\bibinfo {year}
  {2002})}\BibitemShut {NoStop}%
\bibitem [{\citenamefont {Hugel}\ \emph {et~al.}(2002)\citenamefont {Hugel},
  \citenamefont {Holland}, \citenamefont {Cattani}, \citenamefont {Moroder},
  \citenamefont {Seitz},\ and\ \citenamefont {Gaub}}]{Hugel}%
  \BibitemOpen
  \bibfield  {author} {\bibinfo {author} {\bibfnamefont {T.}~\bibnamefont
  {Hugel}}, \bibinfo {author} {\bibfnamefont {N.~B.}\ \bibnamefont {Holland}},
  \bibinfo {author} {\bibfnamefont {A.}~\bibnamefont {Cattani}}, \bibinfo
  {author} {\bibfnamefont {L.}~\bibnamefont {Moroder}}, \bibinfo {author}
  {\bibfnamefont {M.}~\bibnamefont {Seitz}},\ and\ \bibinfo {author}
  {\bibfnamefont {H.~E.}\ \bibnamefont {Gaub}},\ }\bibfield  {title} {\bibinfo
  {title} {{Single-Molecule Optomechanical Cycle}},\ }\href
  {https://doi.org/10.1126/science.1069856} {\bibfield  {journal} {\bibinfo
  {journal} {Science}\ }\textbf {\bibinfo {volume} {296}},\ \bibinfo {pages}
  {1103} (\bibinfo {year} {2002})}\BibitemShut {NoStop}%
\bibitem [{\citenamefont {Roßnagel}\ \emph {et~al.}(2016)\citenamefont
  {Roßnagel}, \citenamefont {Dawkins}, \citenamefont {Tolazzi}, \citenamefont
  {Abah}, \citenamefont {Lutz}, \citenamefont {Schmidt-Kaler},\ and\
  \citenamefont {Singer}}]{Rossnagel}%
  \BibitemOpen
  \bibfield  {author} {\bibinfo {author} {\bibfnamefont {J.}~\bibnamefont
  {Roßnagel}}, \bibinfo {author} {\bibfnamefont {S.~T.}\ \bibnamefont
  {Dawkins}}, \bibinfo {author} {\bibfnamefont {K.~N.}\ \bibnamefont
  {Tolazzi}}, \bibinfo {author} {\bibfnamefont {O.}~\bibnamefont {Abah}},
  \bibinfo {author} {\bibfnamefont {E.}~\bibnamefont {Lutz}}, \bibinfo {author}
  {\bibfnamefont {F.}~\bibnamefont {Schmidt-Kaler}},\ and\ \bibinfo {author}
  {\bibfnamefont {K.}~\bibnamefont {Singer}},\ }\bibfield  {title} {\bibinfo
  {title} {A single-atom heat engine},\ }\href
  {https://doi.org/10.1126/science.aad6320} {\bibfield  {journal} {\bibinfo
  {journal} {Science}\ }\textbf {\bibinfo {volume} {352}},\ \bibinfo {pages}
  {325} (\bibinfo {year} {2016})}\BibitemShut {NoStop}%
\bibitem [{\citenamefont {Kosloff}\ and\ \citenamefont
  {Rezek}(2017)}]{Kosloff2017}%
  \BibitemOpen
  \bibfield  {author} {\bibinfo {author} {\bibfnamefont {R.}~\bibnamefont
  {Kosloff}}\ and\ \bibinfo {author} {\bibfnamefont {Y.}~\bibnamefont
  {Rezek}},\ }\bibfield  {title} {\bibinfo {title} {{The Quantum Harmonic Otto
  Cycle}},\ }\href {https://www.mdpi.com/1099-4300/19/4/136} {\bibfield
  {journal} {\bibinfo  {journal} {Entropy}\ }\textbf {\bibinfo {volume} {19}},\
  \bibinfo {pages} {136} (\bibinfo {year} {2017})}\BibitemShut {NoStop}%
\bibitem [{\citenamefont {Goswami}\ and\ \citenamefont
  {Harbola}(2013)}]{Goswami}%
  \BibitemOpen
  \bibfield  {author} {\bibinfo {author} {\bibfnamefont {H.~P.}\ \bibnamefont
  {Goswami}}\ and\ \bibinfo {author} {\bibfnamefont {U.}~\bibnamefont
  {Harbola}},\ }\bibfield  {title} {\bibinfo {title} {Thermodynamics of quantum
  heat engines},\ }\href {https://doi.org/10.1103/PhysRevA.88.013842}
  {\bibfield  {journal} {\bibinfo  {journal} {Phys. Rev. A}\ }\textbf {\bibinfo
  {volume} {88}},\ \bibinfo {pages} {013842} (\bibinfo {year}
  {2013})}\BibitemShut {NoStop}%
\bibitem [{\citenamefont {Newman}\ \emph {et~al.}(2017)\citenamefont {Newman},
  \citenamefont {Mintert},\ and\ \citenamefont {Nazir}}]{Newman}%
  \BibitemOpen
  \bibfield  {author} {\bibinfo {author} {\bibfnamefont {D.}~\bibnamefont
  {Newman}}, \bibinfo {author} {\bibfnamefont {F.}~\bibnamefont {Mintert}},\
  and\ \bibinfo {author} {\bibfnamefont {A.}~\bibnamefont {Nazir}},\ }\bibfield
   {title} {\bibinfo {title} {Performance of a quantum heat engine at strong
  reservoir coupling},\ }\href {https://doi.org/10.1103/PhysRevE.95.032139}
  {\bibfield  {journal} {\bibinfo  {journal} {Phys. Rev. E}\ }\textbf {\bibinfo
  {volume} {95}},\ \bibinfo {pages} {032139} (\bibinfo {year}
  {2017})}\BibitemShut {NoStop}%
\bibitem [{\citenamefont {Ono}\ \emph {et~al.}(2020)\citenamefont {Ono},
  \citenamefont {Shevchenko}, \citenamefont {Mori}, \citenamefont {Moriyama},\
  and\ \citenamefont {Nori}}]{Ono}%
  \BibitemOpen
  \bibfield  {author} {\bibinfo {author} {\bibfnamefont {K.}~\bibnamefont
  {Ono}}, \bibinfo {author} {\bibfnamefont {S.~N.}\ \bibnamefont {Shevchenko}},
  \bibinfo {author} {\bibfnamefont {T.}~\bibnamefont {Mori}}, \bibinfo {author}
  {\bibfnamefont {S.}~\bibnamefont {Moriyama}},\ and\ \bibinfo {author}
  {\bibfnamefont {F.}~\bibnamefont {Nori}},\ }\bibfield  {title} {\bibinfo
  {title} {{Analog of a Quantum Heat Engine Using a Single-Spin Qubit}},\
  }\href {https://doi.org/10.1103/PhysRevLett.125.166802} {\bibfield  {journal}
  {\bibinfo  {journal} {Phys. Rev. Lett.}\ }\textbf {\bibinfo {volume} {125}},\
  \bibinfo {pages} {166802} (\bibinfo {year} {2020})}\BibitemShut {NoStop}%
\bibitem [{\citenamefont {Jaramillo}\ \emph {et~al.}(2016)\citenamefont
  {Jaramillo}, \citenamefont {Beau},\ and\ \citenamefont {del
  Campo}}]{Jaramillo_2016}%
  \BibitemOpen
  \bibfield  {author} {\bibinfo {author} {\bibfnamefont {J.}~\bibnamefont
  {Jaramillo}}, \bibinfo {author} {\bibfnamefont {M.}~\bibnamefont {Beau}},\
  and\ \bibinfo {author} {\bibfnamefont {A.}~\bibnamefont {del Campo}},\
  }\bibfield  {title} {\bibinfo {title} {Quantum supremacy of many-particle
  thermal machines},\ }\href {https://doi.org/10.1088/1367-2630/18/7/075019}
  {\bibfield  {journal} {\bibinfo  {journal} {New Journal of Physics}\ }\textbf
  {\bibinfo {volume} {18}},\ \bibinfo {pages} {075019} (\bibinfo {year}
  {2016})}\BibitemShut {NoStop}%
\bibitem [{\citenamefont {Chen}\ \emph {et~al.}(2019)\citenamefont {Chen},
  \citenamefont {Watanabe}, \citenamefont {Yu}, \citenamefont {Guan},\ and\
  \citenamefont {del Campo}}]{Chen_2019}%
  \BibitemOpen
  \bibfield  {author} {\bibinfo {author} {\bibfnamefont {Y.-Y.}\ \bibnamefont
  {Chen}}, \bibinfo {author} {\bibfnamefont {G.}~\bibnamefont {Watanabe}},
  \bibinfo {author} {\bibfnamefont {Y.-C.}\ \bibnamefont {Yu}}, \bibinfo
  {author} {\bibfnamefont {X.-W.}\ \bibnamefont {Guan}},\ and\ \bibinfo
  {author} {\bibfnamefont {A.}~\bibnamefont {del Campo}},\ }\bibfield  {title}
  {\bibinfo {title} {An interaction-driven many-particle quantum heat engine
  and its universal behavior},\ }\href
  {https://doi.org/10.1038/s41534-019-0204-5} {\bibfield  {journal} {\bibinfo
  {journal} {npj Quantum Inf}\ }\textbf {\bibinfo {volume} {5}},\ \bibinfo
  {pages} {88} (\bibinfo {year} {2019})}\BibitemShut {NoStop}%
\bibitem [{\citenamefont {Carollo}\ \emph {et~al.}(2020)\citenamefont
  {Carollo}, \citenamefont {Gambetta}, \citenamefont {Brandner}, \citenamefont
  {Garrahan},\ and\ \citenamefont {Lesanovsky}}]{Carollo_2020}%
  \BibitemOpen
  \bibfield  {author} {\bibinfo {author} {\bibfnamefont {F.}~\bibnamefont
  {Carollo}}, \bibinfo {author} {\bibfnamefont {F.~M.}\ \bibnamefont
  {Gambetta}}, \bibinfo {author} {\bibfnamefont {K.}~\bibnamefont {Brandner}},
  \bibinfo {author} {\bibfnamefont {J.~P.}\ \bibnamefont {Garrahan}},\ and\
  \bibinfo {author} {\bibfnamefont {I.}~\bibnamefont {Lesanovsky}},\ }\bibfield
   {title} {\bibinfo {title} {{Nonequilibrium Quantum Many-Body Rydberg Atom
  Engine}},\ }\href {https://doi.org/10.1103/PhysRevLett.124.170602} {\bibfield
   {journal} {\bibinfo  {journal} {Phys. Rev. Lett.}\ }\textbf {\bibinfo
  {volume} {124}},\ \bibinfo {pages} {170602} (\bibinfo {year}
  {2020})}\BibitemShut {NoStop}%
\bibitem [{\citenamefont {Myers}\ \emph {et~al.}(2021)\citenamefont {Myers},
  \citenamefont {McCready},\ and\ \citenamefont {Deffner}}]{Myers}%
  \BibitemOpen
  \bibfield  {author} {\bibinfo {author} {\bibfnamefont {N.~M.}\ \bibnamefont
  {Myers}}, \bibinfo {author} {\bibfnamefont {J.}~\bibnamefont {McCready}},\
  and\ \bibinfo {author} {\bibfnamefont {S.}~\bibnamefont {Deffner}},\
  }\bibfield  {title} {\bibinfo {title} {{Quantum Heat Engines with Singular
  Interactions}},\ }\bibfield  {journal} {\bibinfo  {journal} {Symmetry}\
  }\textbf {\bibinfo {volume} {13}},\ \href
  {https://doi.org/10.3390/sym13060978} {10.3390/sym13060978} (\bibinfo {year}
  {2021})\BibitemShut {NoStop}%
\bibitem [{\citenamefont {Fogarty}\ and\ \citenamefont
  {Busch}(2020)}]{Fogarty_2020}%
  \BibitemOpen
  \bibfield  {author} {\bibinfo {author} {\bibfnamefont {T.}~\bibnamefont
  {Fogarty}}\ and\ \bibinfo {author} {\bibfnamefont {T.}~\bibnamefont
  {Busch}},\ }\bibfield  {title} {\bibinfo {title} {{A many-body heat engine at
  criticality}},\ }\href {https://doi.org/10.1088/2058-9565/abbc63} {\bibfield
  {journal} {\bibinfo  {journal} {Quantum Science and Technology}\ }\textbf
  {\bibinfo {volume} {6}},\ \bibinfo {pages} {015003} (\bibinfo {year}
  {2020})}\BibitemShut {NoStop}%
\bibitem [{\citenamefont {Yunger~Halpern}\ \emph {et~al.}(2019)\citenamefont
  {Yunger~Halpern}, \citenamefont {White}, \citenamefont {Gopalakrishnan},\
  and\ \citenamefont {Refael}}]{Halpern}%
  \BibitemOpen
  \bibfield  {author} {\bibinfo {author} {\bibfnamefont {N.}~\bibnamefont
  {Yunger~Halpern}}, \bibinfo {author} {\bibfnamefont {C.~D.}\ \bibnamefont
  {White}}, \bibinfo {author} {\bibfnamefont {S.}~\bibnamefont
  {Gopalakrishnan}},\ and\ \bibinfo {author} {\bibfnamefont {G.}~\bibnamefont
  {Refael}},\ }\bibfield  {title} {\bibinfo {title} {Quantum engine based on
  many-body localization},\ }\href {https://doi.org/10.1103/PhysRevB.99.024203}
  {\bibfield  {journal} {\bibinfo  {journal} {Phys. Rev. B}\ }\textbf {\bibinfo
  {volume} {99}},\ \bibinfo {pages} {024203} (\bibinfo {year}
  {2019})}\BibitemShut {NoStop}%
\bibitem [{\citenamefont {Uzdin}(2016)}]{Uzdin}%
  \BibitemOpen
  \bibfield  {author} {\bibinfo {author} {\bibfnamefont {R.}~\bibnamefont
  {Uzdin}},\ }\bibfield  {title} {\bibinfo {title} {{Coherence-Induced
  Reversibility and Collective Operation of Quantum Heat Machines via Coherence
  Recycling}},\ }\href {https://doi.org/10.1103/PhysRevApplied.6.024004}
  {\bibfield  {journal} {\bibinfo  {journal} {Phys. Rev. Applied}\ }\textbf
  {\bibinfo {volume} {6}},\ \bibinfo {pages} {024004} (\bibinfo {year}
  {2016})}\BibitemShut {NoStop}%
\bibitem [{\citenamefont {Vroylandt}\ \emph {et~al.}(2017)\citenamefont
  {Vroylandt}, \citenamefont {Esposito},\ and\ \citenamefont
  {Verley}}]{Vroylandt_2017}%
  \BibitemOpen
  \bibfield  {author} {\bibinfo {author} {\bibfnamefont {H.}~\bibnamefont
  {Vroylandt}}, \bibinfo {author} {\bibfnamefont {M.}~\bibnamefont
  {Esposito}},\ and\ \bibinfo {author} {\bibfnamefont {G.}~\bibnamefont
  {Verley}},\ }\bibfield  {title} {\bibinfo {title} {Collective effects
  enhancing power and efficiency},\ }\href
  {https://doi.org/10.1209/0295-5075/120/30009} {\bibfield  {journal} {\bibinfo
   {journal} {{EPL} (Europhysics Letters)}\ }\textbf {\bibinfo {volume}
  {120}},\ \bibinfo {pages} {30009} (\bibinfo {year} {2017})}\BibitemShut
  {NoStop}%
\bibitem [{\citenamefont {Niedenzu}\ and\ \citenamefont
  {Kurizki}(2018)}]{Niedenzu_2018}%
  \BibitemOpen
  \bibfield  {author} {\bibinfo {author} {\bibfnamefont {W.}~\bibnamefont
  {Niedenzu}}\ and\ \bibinfo {author} {\bibfnamefont {G.}~\bibnamefont
  {Kurizki}},\ }\bibfield  {title} {\bibinfo {title} {Cooperative many-body
  enhancement of quantum thermal machine power},\ }\href
  {https://doi.org/10.1088/1367-2630/aaed55} {\bibfield  {journal} {\bibinfo
  {journal} {New Journal of Physics}\ }\textbf {\bibinfo {volume} {20}},\
  \bibinfo {pages} {113038} (\bibinfo {year} {2018})}\BibitemShut {NoStop}%
\bibitem [{\citenamefont {Myers}\ \emph {et~al.}(2022)\citenamefont {Myers},
  \citenamefont {Peña}, \citenamefont {Negrete}, \citenamefont {Vargas},
  \citenamefont {Chiara},\ and\ \citenamefont {Deffner}}]{Myers_2022}%
  \BibitemOpen
  \bibfield  {author} {\bibinfo {author} {\bibfnamefont {N.~M.}\ \bibnamefont
  {Myers}}, \bibinfo {author} {\bibfnamefont {F.~J.}\ \bibnamefont {Peña}},
  \bibinfo {author} {\bibfnamefont {O.}~\bibnamefont {Negrete}}, \bibinfo
  {author} {\bibfnamefont {P.}~\bibnamefont {Vargas}}, \bibinfo {author}
  {\bibfnamefont {G.~D.}\ \bibnamefont {Chiara}},\ and\ \bibinfo {author}
  {\bibfnamefont {S.}~\bibnamefont {Deffner}},\ }\bibfield  {title} {\bibinfo
  {title} {Boosting engine performance with bose–einstein condensation},\
  }\href {https://doi.org/10.1088/1367-2630/ac47cc} {\bibfield  {journal}
  {\bibinfo  {journal} {New Journal of Physics}\ }\textbf {\bibinfo {volume}
  {24}},\ \bibinfo {pages} {025001} (\bibinfo {year} {2022})}\BibitemShut
  {NoStop}%
\bibitem [{\citenamefont {Chen}\ \emph {et~al.}(2018)\citenamefont {Chen},
  \citenamefont {Dong},\ and\ \citenamefont {Sun}}]{Chen}%
  \BibitemOpen
  \bibfield  {author} {\bibinfo {author} {\bibfnamefont {J.}~\bibnamefont
  {Chen}}, \bibinfo {author} {\bibfnamefont {H.}~\bibnamefont {Dong}},\ and\
  \bibinfo {author} {\bibfnamefont {C.-P.}\ \bibnamefont {Sun}},\ }\bibfield
  {title} {\bibinfo {title} {{Bose-Fermi duality in a quantum Otto heat engine
  with trapped repulsive bosons}},\ }\href
  {https://doi.org/10.1103/PhysRevE.98.062119} {\bibfield  {journal} {\bibinfo
  {journal} {Phys. Rev. E}\ }\textbf {\bibinfo {volume} {98}},\ \bibinfo
  {pages} {062119} (\bibinfo {year} {2018})}\BibitemShut {NoStop}%
\bibitem [{\citenamefont {Keller}\ \emph {et~al.}(2020)\citenamefont {Keller},
  \citenamefont {Fogarty}, \citenamefont {Li},\ and\ \citenamefont
  {Busch}}]{Keller_2020}%
  \BibitemOpen
  \bibfield  {author} {\bibinfo {author} {\bibfnamefont {T.}~\bibnamefont
  {Keller}}, \bibinfo {author} {\bibfnamefont {T.}~\bibnamefont {Fogarty}},
  \bibinfo {author} {\bibfnamefont {J.}~\bibnamefont {Li}},\ and\ \bibinfo
  {author} {\bibfnamefont {T.}~\bibnamefont {Busch}},\ }\bibfield  {title}
  {\bibinfo {title} {{Feshbach engine in the Thomas-Fermi regime}},\ }\href
  {https://doi.org/10.1103/PhysRevResearch.2.033335} {\bibfield  {journal}
  {\bibinfo  {journal} {Phys. Rev. Research}\ }\textbf {\bibinfo {volume}
  {2}},\ \bibinfo {pages} {033335} (\bibinfo {year} {2020})}\BibitemShut
  {NoStop}%
\bibitem [{\citenamefont {Hartmann}\ \emph {et~al.}(2020)\citenamefont
  {Hartmann}, \citenamefont {Mukherjee}, \citenamefont {Niedenzu},\ and\
  \citenamefont {Lechner}}]{Hartmann_2020}%
  \BibitemOpen
  \bibfield  {author} {\bibinfo {author} {\bibfnamefont {A.}~\bibnamefont
  {Hartmann}}, \bibinfo {author} {\bibfnamefont {V.}~\bibnamefont {Mukherjee}},
  \bibinfo {author} {\bibfnamefont {W.}~\bibnamefont {Niedenzu}},\ and\
  \bibinfo {author} {\bibfnamefont {W.}~\bibnamefont {Lechner}},\ }\bibfield
  {title} {\bibinfo {title} {Many-body quantum heat engines with shortcuts to
  adiabaticity},\ }\href {https://doi.org/10.1103/PhysRevResearch.2.023145}
  {\bibfield  {journal} {\bibinfo  {journal} {Phys. Rev. Research}\ }\textbf
  {\bibinfo {volume} {2}},\ \bibinfo {pages} {023145} (\bibinfo {year}
  {2020})}\BibitemShut {NoStop}%
\bibitem [{\citenamefont {Beau}\ \emph {et~al.}(2016)\citenamefont {Beau},
  \citenamefont {Jaramillo},\ and\ \citenamefont {Del~Campo}}]{Beau_2016}%
  \BibitemOpen
  \bibfield  {author} {\bibinfo {author} {\bibfnamefont {M.}~\bibnamefont
  {Beau}}, \bibinfo {author} {\bibfnamefont {J.}~\bibnamefont {Jaramillo}},\
  and\ \bibinfo {author} {\bibfnamefont {A.}~\bibnamefont {Del~Campo}},\
  }\bibfield  {title} {\bibinfo {title} {{Scaling-Up Quantum Heat Engines
  Efficiently via Shortcuts to Adiabaticity}},\ }\href
  {https://www.mdpi.com/1099-4300/18/5/168} {\bibfield  {journal} {\bibinfo
  {journal} {Entropy}\ }\textbf {\bibinfo {volume} {18}},\ \bibinfo {pages}
  {168} (\bibinfo {year} {2016})}\BibitemShut {NoStop}%
\bibitem [{\citenamefont {Deng}\ \emph {et~al.}(2018)\citenamefont {Deng},
  \citenamefont {Chenu}, \citenamefont {Diao}, \citenamefont {Li},
  \citenamefont {Yu}, \citenamefont {Coulamy}, \citenamefont {del Campo},\ and\
  \citenamefont {Wu}}]{Deng_2018}%
  \BibitemOpen
  \bibfield  {author} {\bibinfo {author} {\bibfnamefont {S.}~\bibnamefont
  {Deng}}, \bibinfo {author} {\bibfnamefont {A.}~\bibnamefont {Chenu}},
  \bibinfo {author} {\bibfnamefont {P.}~\bibnamefont {Diao}}, \bibinfo {author}
  {\bibfnamefont {F.}~\bibnamefont {Li}}, \bibinfo {author} {\bibfnamefont
  {S.}~\bibnamefont {Yu}}, \bibinfo {author} {\bibfnamefont {I.}~\bibnamefont
  {Coulamy}}, \bibinfo {author} {\bibfnamefont {A.}~\bibnamefont {del Campo}},\
  and\ \bibinfo {author} {\bibfnamefont {H.}~\bibnamefont {Wu}},\ }\bibfield
  {title} {\bibinfo {title} {Superadiabatic quantum friction suppression in
  finite-time thermodynamics},\ }\href {https://doi.org/10.1126/sciadv.aar5909}
  {\bibfield  {journal} {\bibinfo  {journal} {Science Advances}\ }\textbf
  {\bibinfo {volume} {4}},\ \bibinfo {pages} {5909} (\bibinfo {year}
  {2018})}\BibitemShut {NoStop}%
\bibitem [{\citenamefont {Li}\ \emph {et~al.}(2018)\citenamefont {Li},
  \citenamefont {Fogarty}, \citenamefont {Campbell}, \citenamefont {Chen},\
  and\ \citenamefont {Busch}}]{Li_2018}%
  \BibitemOpen
  \bibfield  {author} {\bibinfo {author} {\bibfnamefont {J.}~\bibnamefont
  {Li}}, \bibinfo {author} {\bibfnamefont {T.}~\bibnamefont {Fogarty}},
  \bibinfo {author} {\bibfnamefont {S.}~\bibnamefont {Campbell}}, \bibinfo
  {author} {\bibfnamefont {X.}~\bibnamefont {Chen}},\ and\ \bibinfo {author}
  {\bibfnamefont {T.}~\bibnamefont {Busch}},\ }\bibfield  {title} {\bibinfo
  {title} {{An efficient nonlinear Feshbach engine}},\ }\href
  {https://doi.org/10.1088/1367-2630/aa9cd8} {\bibfield  {journal} {\bibinfo
  {journal} {New Journal of Physics}\ }\textbf {\bibinfo {volume} {20}},\
  \bibinfo {pages} {015005} (\bibinfo {year} {2018})}\BibitemShut {NoStop}%
\bibitem [{\citenamefont {Fogarty}\ \emph {et~al.}(2019)\citenamefont
  {Fogarty}, \citenamefont {Ruks}, \citenamefont {Li},\ and\ \citenamefont
  {Busch}}]{Mossy}%
  \BibitemOpen
  \bibfield  {author} {\bibinfo {author} {\bibfnamefont {T.}~\bibnamefont
  {Fogarty}}, \bibinfo {author} {\bibfnamefont {L.}~\bibnamefont {Ruks}},
  \bibinfo {author} {\bibfnamefont {J.}~\bibnamefont {Li}},\ and\ \bibinfo
  {author} {\bibfnamefont {T.}~\bibnamefont {Busch}},\ }\bibfield  {title}
  {\bibinfo {title} {{Fast control of interactions in an ultracold two atom
  system: Managing correlations and irreversibility}},\ }\href
  {https://doi.org/10.21468/SciPostPhys.6.2.021} {\bibfield  {journal}
  {\bibinfo  {journal} {SciPost Phys.}\ }\textbf {\bibinfo {volume} {6}},\
  \bibinfo {pages} {21} (\bibinfo {year} {2019})}\BibitemShut {NoStop}%
\bibitem [{\citenamefont {Mikkelsen}\ \emph {et~al.}(2022)\citenamefont
  {Mikkelsen}, \citenamefont {Fogarty},\ and\ \citenamefont
  {Busch}}]{Mikkelsen}%
  \BibitemOpen
  \bibfield  {author} {\bibinfo {author} {\bibfnamefont {M.}~\bibnamefont
  {Mikkelsen}}, \bibinfo {author} {\bibfnamefont {T.}~\bibnamefont {Fogarty}},\
  and\ \bibinfo {author} {\bibfnamefont {T.}~\bibnamefont {Busch}},\ }\bibfield
   {title} {\bibinfo {title} {Connecting scrambling and work statistics for
  short-range interactions in the harmonic oscillator},\ }\href
  {https://doi.org/10.1103/PhysRevLett.128.070605} {\bibfield  {journal}
  {\bibinfo  {journal} {Phys. Rev. Lett.}\ }\textbf {\bibinfo {volume} {128}},\
  \bibinfo {pages} {070605} (\bibinfo {year} {2022})}\BibitemShut {NoStop}%
\bibitem [{\citenamefont {Busch}\ \emph {et~al.}(1998)\citenamefont {Busch},
  \citenamefont {Englert}, \citenamefont {Rzażewski},\ and\ \citenamefont
  {Wilkens}}]{Busch98}%
  \BibitemOpen
  \bibfield  {author} {\bibinfo {author} {\bibfnamefont {T.}~\bibnamefont
  {Busch}}, \bibinfo {author} {\bibfnamefont {B.-G.}\ \bibnamefont {Englert}},
  \bibinfo {author} {\bibfnamefont {K.}~\bibnamefont {Rzażewski}},\ and\
  \bibinfo {author} {\bibfnamefont {M.}~\bibnamefont {Wilkens}},\ }\bibfield
  {title} {\bibinfo {title} {{Two Cold Atoms in a Harmonic Trap}},\ }\href
  {https://doi.org/10.1023/A:1018705520999} {\bibfield  {journal} {\bibinfo
  {journal} {Foundations of Physics}\ }\textbf {\bibinfo {volume} {28}},\
  \bibinfo {pages} {549} (\bibinfo {year} {1998})}\BibitemShut {NoStop}%
\bibitem [{\citenamefont {Wei{\ss}e}\ and\ \citenamefont
  {Fehske}(2008)}]{Weisse_2008}%
  \BibitemOpen
  \bibfield  {author} {\bibinfo {author} {\bibfnamefont {A.}~\bibnamefont
  {Wei{\ss}e}}\ and\ \bibinfo {author} {\bibfnamefont {H.}~\bibnamefont
  {Fehske}},\ }\bibinfo {title} {{Exact Diagonalization Techniques}},\ in\
  \href {https://doi.org/10.1007/978-3-540-74686-7_18} {\emph {\bibinfo
  {booktitle} {Computational Many-Particle Physics}}},\ \bibinfo {editor}
  {edited by\ \bibinfo {editor} {\bibfnamefont {H.}~\bibnamefont {Fehske}},
  \bibinfo {editor} {\bibfnamefont {R.}~\bibnamefont {Schneider}},\ and\
  \bibinfo {editor} {\bibfnamefont {A.}~\bibnamefont {Wei{\ss}e}}}\ (\bibinfo
  {publisher} {Springer Berlin Heidelberg},\ \bibinfo {address} {Berlin,
  Heidelberg},\ \bibinfo {year} {2008})\ pp.\ \bibinfo {pages}
  {529--544}\BibitemShut {NoStop}%
\bibitem [{\citenamefont {Na}\ and\ \citenamefont {Marsiglio}(2017)}]{Na_2017}%
  \BibitemOpen
  \bibfield  {author} {\bibinfo {author} {\bibfnamefont {M.}~\bibnamefont
  {Na}}\ and\ \bibinfo {author} {\bibfnamefont {F.}~\bibnamefont {Marsiglio}},\
  }\bibfield  {title} {\bibinfo {title} {Two and three particles interacting in
  a one-dimensional trap},\ }\href {https://doi.org/10.1119/1.4985063}
  {\bibfield  {journal} {\bibinfo  {journal} {American Journal of Physics}\
  }\textbf {\bibinfo {volume} {85}},\ \bibinfo {pages} {769} (\bibinfo {year}
  {2017})}\BibitemShut {NoStop}%
\bibitem [{\citenamefont {Girardeau}(1960)}]{Girardeau}%
  \BibitemOpen
  \bibfield  {author} {\bibinfo {author} {\bibfnamefont {M.}~\bibnamefont
  {Girardeau}},\ }\bibfield  {title} {\bibinfo {title} {{Relationship between
  Systems of Impenetrable Bosons and Fermions in One Dimension}},\ }\href
  {https://doi.org/10.1063/1.1703687} {\bibfield  {journal} {\bibinfo
  {journal} {Journal of Mathematical Physics}\ }\textbf {\bibinfo {volume}
  {1}},\ \bibinfo {pages} {516} (\bibinfo {year} {1960})}\BibitemShut {NoStop}%
\bibitem [{\citenamefont {de~Oliveira}\ and\ \citenamefont
  {Jonathan}(2021)}]{Oliveira}%
  \BibitemOpen
  \bibfield  {author} {\bibinfo {author} {\bibfnamefont {T.~R.}\ \bibnamefont
  {de~Oliveira}}\ and\ \bibinfo {author} {\bibfnamefont {D.}~\bibnamefont
  {Jonathan}},\ }\bibfield  {title} {\bibinfo {title} {Efficiency gain and
  bidirectional operation of quantum engines with decoupled internal levels},\
  }\href {https://doi.org/10.1103/PhysRevE.104.044133} {\bibfield  {journal}
  {\bibinfo  {journal} {Phys. Rev. E}\ }\textbf {\bibinfo {volume} {104}},\
  \bibinfo {pages} {044133} (\bibinfo {year} {2021})}\BibitemShut {NoStop}%
\bibitem [{\citenamefont {Curzon}\ and\ \citenamefont {Ahlborn}(1975)}]{CA}%
  \BibitemOpen
  \bibfield  {author} {\bibinfo {author} {\bibfnamefont {F.~L.}\ \bibnamefont
  {Curzon}}\ and\ \bibinfo {author} {\bibfnamefont {B.}~\bibnamefont
  {Ahlborn}},\ }\bibfield  {title} {\bibinfo {title} {{Efficiency of a Carnot
  engine at maximum power output}},\ }\href {https://doi.org/10.1119/1.10023}
  {\bibfield  {journal} {\bibinfo  {journal} {American Journal of Physics}\
  }\textbf {\bibinfo {volume} {43}},\ \bibinfo {pages} {22} (\bibinfo {year}
  {1975})}\BibitemShut {NoStop}%
\bibitem [{\citenamefont {Leff}(1987)}]{Leff}%
  \BibitemOpen
  \bibfield  {author} {\bibinfo {author} {\bibfnamefont {H.~S.}\ \bibnamefont
  {Leff}},\ }\bibfield  {title} {\bibinfo {title} {{Thermal efficiency at
  maximum work output: New results for old heat engines}},\ }\href
  {https://doi.org/10.1119/1.15071} {\bibfield  {journal} {\bibinfo  {journal}
  {American Journal of Physics}\ }\textbf {\bibinfo {volume} {55}},\ \bibinfo
  {pages} {602} (\bibinfo {year} {1987})}\BibitemShut {NoStop}%
\bibitem [{\citenamefont {Rezek}\ and\ \citenamefont {Kosloff}(2006)}]{Rezek}%
  \BibitemOpen
  \bibfield  {author} {\bibinfo {author} {\bibfnamefont {Y.}~\bibnamefont
  {Rezek}}\ and\ \bibinfo {author} {\bibfnamefont {R.}~\bibnamefont
  {Kosloff}},\ }\bibfield  {title} {\bibinfo {title} {Irreversible performance
  of a quantum harmonic heat engine},\ }\href
  {https://doi.org/10.1088/1367-2630/8/5/083} {\bibfield  {journal} {\bibinfo
  {journal} {New Journal of Physics}\ }\textbf {\bibinfo {volume} {8}},\
  \bibinfo {pages} {83} (\bibinfo {year} {2006})}\BibitemShut {NoStop}%
\bibitem [{\citenamefont {Alecce}\ \emph {et~al.}(2015)\citenamefont {Alecce},
  \citenamefont {Galve}, \citenamefont {Gullo}, \citenamefont {Dell'Anna},
  \citenamefont {Plastina},\ and\ \citenamefont {Zambrini}}]{Alecce_2015}%
  \BibitemOpen
  \bibfield  {author} {\bibinfo {author} {\bibfnamefont {A.}~\bibnamefont
  {Alecce}}, \bibinfo {author} {\bibfnamefont {F.}~\bibnamefont {Galve}},
  \bibinfo {author} {\bibfnamefont {N.~L.}\ \bibnamefont {Gullo}}, \bibinfo
  {author} {\bibfnamefont {L.}~\bibnamefont {Dell'Anna}}, \bibinfo {author}
  {\bibfnamefont {F.}~\bibnamefont {Plastina}},\ and\ \bibinfo {author}
  {\bibfnamefont {R.}~\bibnamefont {Zambrini}},\ }\bibfield  {title} {\bibinfo
  {title} {{Quantum Otto cycle with inner friction: finite-time and disorder
  effects}},\ }\href {https://doi.org/10.1088/1367-2630/17/7/075007} {\bibfield
   {journal} {\bibinfo  {journal} {New Journal of Physics}\ }\textbf {\bibinfo
  {volume} {17}},\ \bibinfo {pages} {075007} (\bibinfo {year}
  {2015})}\BibitemShut {NoStop}%
\bibitem [{\citenamefont {Feldmann}\ and\ \citenamefont
  {Kosloff}(2000)}]{Feldmann}%
  \BibitemOpen
  \bibfield  {author} {\bibinfo {author} {\bibfnamefont {T.}~\bibnamefont
  {Feldmann}}\ and\ \bibinfo {author} {\bibfnamefont {R.}~\bibnamefont
  {Kosloff}},\ }\bibfield  {title} {\bibinfo {title} {Performance of discrete
  heat engines and heat pumps in finite time},\ }\href
  {https://doi.org/10.1103/PhysRevE.61.4774} {\bibfield  {journal} {\bibinfo
  {journal} {Phys. Rev. E}\ }\textbf {\bibinfo {volume} {61}},\ \bibinfo
  {pages} {4774} (\bibinfo {year} {2000})}\BibitemShut {NoStop}%
\bibitem [{\citenamefont {Feldmann}\ and\ \citenamefont
  {Kosloff}(2003)}]{Feldmann_2003}%
  \BibitemOpen
  \bibfield  {author} {\bibinfo {author} {\bibfnamefont {T.}~\bibnamefont
  {Feldmann}}\ and\ \bibinfo {author} {\bibfnamefont {R.}~\bibnamefont
  {Kosloff}},\ }\bibfield  {title} {\bibinfo {title} {{Quantum four-stroke heat
  engine: Thermodynamic observables in a model with intrinsic friction}},\
  }\href {https://doi.org/10.1103/PhysRevE.68.016101} {\bibfield  {journal}
  {\bibinfo  {journal} {Phys. Rev. E}\ }\textbf {\bibinfo {volume} {68}},\
  \bibinfo {pages} {016101} (\bibinfo {year} {2003})}\BibitemShut {NoStop}%
\bibitem [{\citenamefont {Çakmak}\ \emph {et~al.}(2017)\citenamefont
  {Çakmak}, \citenamefont {Altintas}, \citenamefont {Gençten}, ,\ and\
  \citenamefont {M\"ustecaplıoğlu}}]{Cakmak_2017}%
  \BibitemOpen
  \bibfield  {author} {\bibinfo {author} {\bibfnamefont {S.}~\bibnamefont
  {Çakmak}}, \bibinfo {author} {\bibfnamefont {F.}~\bibnamefont {Altintas}},
  \bibinfo {author} {\bibfnamefont {A.}~\bibnamefont {Gençten}}, ,\ and\
  \bibinfo {author} {\bibfnamefont {O.~E.}\ \bibnamefont
  {M\"ustecaplıoğlu}},\ }\bibfield  {title} {\bibinfo {title} {{Irreversible
  work and internal friction in a quantum Otto cycle of a single arbitrary
  spin}},\ }\href {https://doi.org/10.1140/epjd/e2017-70443-1} {\bibfield
  {journal} {\bibinfo  {journal} {The European Physical Journal D}\ }\textbf
  {\bibinfo {volume} {71}},\ \bibinfo {pages} {016101} (\bibinfo {year}
  {2017})}\BibitemShut {NoStop}%
\bibitem [{\citenamefont {Esposito}\ \emph {et~al.}(2010)\citenamefont
  {Esposito}, \citenamefont {Kawai}, \citenamefont {Lindenberg},\ and\
  \citenamefont {Van~den Broeck}}]{Esposito}%
  \BibitemOpen
  \bibfield  {author} {\bibinfo {author} {\bibfnamefont {M.}~\bibnamefont
  {Esposito}}, \bibinfo {author} {\bibfnamefont {R.}~\bibnamefont {Kawai}},
  \bibinfo {author} {\bibfnamefont {K.}~\bibnamefont {Lindenberg}},\ and\
  \bibinfo {author} {\bibfnamefont {C.}~\bibnamefont {Van~den Broeck}},\
  }\bibfield  {title} {\bibinfo {title} {{Efficiency at Maximum Power of
  Low-Dissipation Carnot Engines}},\ }\href
  {https://doi.org/10.1103/PhysRevLett.105.150603} {\bibfield  {journal}
  {\bibinfo  {journal} {Phys. Rev. Lett.}\ }\textbf {\bibinfo {volume} {105}},\
  \bibinfo {pages} {150603} (\bibinfo {year} {2010})}\BibitemShut {NoStop}%
\bibitem [{\citenamefont {Andresen}(2011)}]{Andresen}%
  \BibitemOpen
  \bibfield  {author} {\bibinfo {author} {\bibfnamefont {B.}~\bibnamefont
  {Andresen}},\ }\bibfield  {title} {\bibinfo {title} {{Current Trends in
  Finite-Time Thermodynamics}},\ }\href
  {https://doi.org/https://doi.org/10.1002/anie.201001411} {\bibfield
  {journal} {\bibinfo  {journal} {Angewandte Chemie International Edition}\
  }\textbf {\bibinfo {volume} {50}},\ \bibinfo {pages} {2690} (\bibinfo {year}
  {2011})}\BibitemShut {NoStop}%
\bibitem [{\citenamefont {Whitney}(2014)}]{Whitney}%
  \BibitemOpen
  \bibfield  {author} {\bibinfo {author} {\bibfnamefont {R.~S.}\ \bibnamefont
  {Whitney}},\ }\bibfield  {title} {\bibinfo {title} {{Most Efficient Quantum
  Thermoelectric at Finite Power Output}},\ }\href
  {https://doi.org/10.1103/PhysRevLett.112.130601} {\bibfield  {journal}
  {\bibinfo  {journal} {Phys. Rev. Lett.}\ }\textbf {\bibinfo {volume} {112}},\
  \bibinfo {pages} {130601} (\bibinfo {year} {2014})}\BibitemShut {NoStop}%
\bibitem [{\citenamefont {Shiraishi}\ \emph {et~al.}(2016)\citenamefont
  {Shiraishi}, \citenamefont {Saito},\ and\ \citenamefont
  {Tasaki}}]{Shiraishi}%
  \BibitemOpen
  \bibfield  {author} {\bibinfo {author} {\bibfnamefont {N.}~\bibnamefont
  {Shiraishi}}, \bibinfo {author} {\bibfnamefont {K.}~\bibnamefont {Saito}},\
  and\ \bibinfo {author} {\bibfnamefont {H.}~\bibnamefont {Tasaki}},\
  }\bibfield  {title} {\bibinfo {title} {{Universal Trade-Off Relation between
  Power and Efficiency for Heat Engines}},\ }\href
  {https://doi.org/10.1103/PhysRevLett.117.190601} {\bibfield  {journal}
  {\bibinfo  {journal} {Phys. Rev. Lett.}\ }\textbf {\bibinfo {volume} {117}},\
  \bibinfo {pages} {190601} (\bibinfo {year} {2016})}\BibitemShut {NoStop}%
\bibitem [{\citenamefont {Lin}\ and\ \citenamefont {Chen}(2003)}]{Lin}%
  \BibitemOpen
  \bibfield  {author} {\bibinfo {author} {\bibfnamefont {B.}~\bibnamefont
  {Lin}}\ and\ \bibinfo {author} {\bibfnamefont {J.}~\bibnamefont {Chen}},\
  }\bibfield  {title} {\bibinfo {title} {Performance analysis of an
  irreversible quantum heat engine working with harmonic oscillators},\ }\href
  {https://doi.org/10.1103/PhysRevE.67.046105} {\bibfield  {journal} {\bibinfo
  {journal} {Phys. Rev. E}\ }\textbf {\bibinfo {volume} {67}},\ \bibinfo
  {pages} {046105} (\bibinfo {year} {2003})}\BibitemShut {NoStop}%
\bibitem [{\citenamefont {Abah}\ and\ \citenamefont {Lutz}(2016)}]{Abah_2016}%
  \BibitemOpen
  \bibfield  {author} {\bibinfo {author} {\bibfnamefont {O.}~\bibnamefont
  {Abah}}\ and\ \bibinfo {author} {\bibfnamefont {E.}~\bibnamefont {Lutz}},\
  }\bibfield  {title} {\bibinfo {title} {{Optimal performance of a quantum Otto
  refrigerator}},\ }\href {https://doi.org/10.1209/0295-5075/113/60002}
  {\bibfield  {journal} {\bibinfo  {journal} {{EPL} (Europhysics Letters)}\
  }\textbf {\bibinfo {volume} {113}},\ \bibinfo {pages} {60002} (\bibinfo
  {year} {2016})}\BibitemShut {NoStop}%
\bibitem [{\citenamefont {Beau}\ \emph {et~al.}(2020)\citenamefont {Beau},
  \citenamefont {Pittman}, \citenamefont {Astrakharchik},\ and\ \citenamefont
  {del Campo}}]{Beau_2020}%
  \BibitemOpen
  \bibfield  {author} {\bibinfo {author} {\bibfnamefont {M.}~\bibnamefont
  {Beau}}, \bibinfo {author} {\bibfnamefont {S.~M.}\ \bibnamefont {Pittman}},
  \bibinfo {author} {\bibfnamefont {G.~E.}\ \bibnamefont {Astrakharchik}},\
  and\ \bibinfo {author} {\bibfnamefont {A.}~\bibnamefont {del Campo}},\
  }\bibfield  {title} {\bibinfo {title} {Exactly solvable system of
  one-dimensional trapped bosons with short- and long-range interactions},\
  }\href {https://doi.org/10.1103/PhysRevLett.125.220602} {\bibfield  {journal}
  {\bibinfo  {journal} {Phys. Rev. Lett.}\ }\textbf {\bibinfo {volume} {125}},\
  \bibinfo {pages} {220602} (\bibinfo {year} {2020})}\BibitemShut {NoStop}%
\end{thebibliography}%

\end{document}